\documentclass[%
 reprint,
preprintnumbers,
nofootinbib,
amsmath,amssymb, aps,
prd,
]{revtex4-2}

\usepackage{graphicx} 
\usepackage{dcolumn}
\usepackage{bm}
\usepackage{enumitem}
\usepackage{xcolor}
    \definecolor{codepurple}{rgb}{0.58,0,0.82}
\usepackage[bottom]{footmisc}
\usepackage{csquotes}
\usepackage{tensor}
\usepackage{mathtools}
\usepackage{physics}
\usepackage{cancel}
\usepackage{makecell}

\usepackage[colorlinks]{hyperref}
\hypersetup{hidelinks, colorlinks = true,
    linkcolor = blue,
    urlcolor = blue,
    citecolor = cyan,
}
\colorlet{linkequation}{codepurple}
\newcommand*{\SavedEqref}{}
\let\SavedEqref\eqref
\renewcommand*{\eqref}[1]{%
  \begingroup
    \hypersetup{
      linkcolor=linkequation,
      linkbordercolor=linkequation,
    }%
    \SavedEqref{#1}%
  \endgroup
}

\begin{document}

\title{Symmetry restoration for TDiff scalar fields}
\author{Darío Jaramillo-Garrido}\email{djaramil@ucm.es}
\author{Antonio L. Maroto}\email{maroto@ucm.es}\author{Prado Martín-Moruno}\email{pradomm@ucm.es}
\affiliation{Departamento de Física Teórica and Instituto de Física de Partículas y del Cosmos (IPARCOS-UCM), Universidad Complutense de Madrid, E-28040 Madrid, Spain}
\date{February 2024}
\begin{abstract}
    We explore the idea of restoring the full diffeomorphism (Diff) invariance in theories with only transverse diffeomorphisms (TDiff) by the introduction of additional fields. In particular, we consider in detail the case of a TDiff invariant scalar field and how Diff symmetry can be restored preserving locality by introducing an additional vector field.  We reobtain the corresponding dynamics and energy-momentum tensor from the covariantized action and analyze the potential and kinetic domination regimes. For the former, the theory describes a cosmological constant-type behaviour, while for the latter we show that the theory can describe an adiabatic perfect fluid whose equation of state and speed of sound is obtained in a straightforward way. Furthermore, the reformulation with the full symmetry allows us to analyze the gravitational properties of the theory beyond those particular regimes. In particular, we find the general expression for the effective speed of sound of the non-adiabatic perfect fluid, which provides us with physically reasonable conditions that should be satisfied by the coupling functions. Finally, we investigate the particular models leading to an adiabatic fluid.
\end{abstract}

\preprint{IPARCOS-UCM-24-013}

\maketitle

\section{Introduction}
Our current best description of gravitational phenomena is, and has been for over a century, the theory of General Relativiy (GR). It is not a theory without weaknesses, however, and although it performs remarkably well in numerous tests on solar system scales, there is reason to believe that it is not the end of the story. For one, it is not a theory which serves to describe gravity in its most extreme regimes, where one expects quantum effects to gain importance, but it is also possible to find problems even while remaining classical. Indeed, the various tensions between theory and observations in cosmology are another hint at the possibility that the theory breaks down at such scales. Sparked by considerations of the sort, together with other theoretical issues such as the cosmological constant problem, modified theories of gravity have been a central object of study in this regard (see e.g. \cite{CANTATA} for a review).

In particular, it proves worthwhile to reconsider the fundamental symmetries involved in GR, namely the diffeomorphism (Diff) invariance of the theory. This amounts to the assertion that the physical equations remain invariant under general coordinate transformations. The study of situations where such a symmetry is broken can in fact be traced back to Einstein himself in 1919 with the introduction of Unimodular Gravity \cite{Einstein}, where the metric determinant is reduced to be a non-dynamical field fixed to the value $g=1$. Unimodular Gravity is perhaps the most well-known example of a theory with broken Diff invariance, being the symmetry group in that case the union of transverse diffeomorphisms (TDiff) and Weyl rescalings (together dubbed WTDiff, see e.g. \cite{Carballo-Rubio} for a review). The equations of motion of the theory are the trace-free Einstein equations (see e.g. \cite{Ellis} for a comprehensive introduction), in which any cosmological constant-type contribution does not gravitate, thus providing an elegant solution to the cosmological constant problem.

In more recent years, interest has grown in theories which present TDiff invariance. Simply put, transverse diffeomorphisms are general coordinate transformations in which the Jacobian determinant is required to be $J=1$. Infinitesimally, if we consider the coordinate transformation $x^\mu \rightarrow \hat{x}^\mu = x^\mu + \xi^\mu(x)$ generated by a vector field $\xi^\mu(x)$, then what we do is require the condition $\partial_\mu \xi^\mu = 0$ (see \cite{Maroto} for a concise introduction to transverse diffeomorphisms). The fact that the Jacobian determinant equals unity means that objects which were tensor densities under Diff become actual tensors under TDiff. In particular, the metric determinant becomes a TDiff scalar field, and this has interesting implications. Indeed, on the one hand, the metric determinant becomes a scalar field to which one may endow dynamics. On the other hand, the invariant volume element we find in an action integral is no longer fixed by the symmetry to be $d\text{vol}= \sqrt{g} \, d^4x$, but can actually take on the more general form $d\text{vol}=f(g)\,d^4x$, with $f(g)$ an arbitrary function of the metric determinant, and this opens up an enormous range of possibilities for novel couplings. Field theories in which the gravitational sector is TDiff invariant are studied in references \cite{Bello-Morales,Pirogov:2009,Pirogov:2011,Pirogov:2005,Pirogov:2015}, where cosmological implications are also discussed. One can also study the consequences of breaking the symmetries in the matter sector. TDiff invariant theories with a scalar field were recently considered in references \cite{Alvarez,Maroto,Jaramillo-Garrido,Alonso-Lopez}. Reference \cite{Maroto} considers general scalar field TDiff theories in cosmological contexts, reference \cite{Jaramillo-Garrido} performs a general study for a scalar field without assuming any background geometry, and reference \cite{Alonso-Lopez} provides a unified description for the dark sector using a particular theory, comparing the results with the latest cosmological observations and datasets.

Now, it is not an uncommon situation in physics to find several equivalent descriptions, or reformulations, of the same theory. Examples include, but are by no means limited to, the well-known equivalence between the Palatini and the metric approach to GR, the scalar-tensor perspective of $f(R)$-gravity (see, for example, \cite{CANTATA}), or the correspondence between the field equations of non linear Ricci-based metric-affine theories of gravity coupled to scalar matter and GR coupled to a different scalar field Lagrangian \cite{Afonso:2018hyj}.
Moreover, ever since the pioneering work of Stueckelberg \cite{Stueckelberg} (see e.g. \cite{Ruegg} for a review), it is well known how a theory may recover (or reveal) its broken gauge symmetry via the introduction of additional fields. With these two ideas in mind, we present in this work an alternative formalism for the treatment of a TDiff theory for a scalar field based on previous work done by Henneaux and Teitelboim \cite{Henneaux}  within the context of Unimodular Gravity (see also Kuchar \cite{Kuchar} for an alternative approach). Following that spirit, we shall in this work show how a TDiff invariant field theory may be equivalently described as a Diff invariant theory with an additional field. We will then work on the particular case of a scalar field theory from both points of view.
This idea of finding an equivalent description with symmetry restoration has been also applied on the different gravitational framework of massive gravity \cite{deRham:2011rn}.

The paper is organized as follows. First of all,  in section \ref{section: The TDiff approach} we review the TDiff approach for the scalar field  and summarize the main results for the potential domination and kinetic domination regimes in sections \ref{subsecIIIA} and \ref{subsecIIIB}. In section \ref{section: Covariantized action} we show how it is possible to reformulate a TDiff invariant field theory in a way which recovers Diff invariance via the introduction of an additional field. Section \ref{section: The covariantized approach} is then devoted to the reformulation of our scalar field theory in a covariantized manner. In section \ref{subsecIVA} we recover the results of the TDiff approach in the potential domination regime. Then, in section \ref{subsecIVB}, we not only re-obtain the results in the kinetic domination regime, but the use of the covariantized approach allows us to study the stability of the adiabatic fluid in a simple way. Moreover, in section \ref{subsecIVC} we argue how the use of a particular approach could lead to the study of different kinetic models considered to be more natural. 
In addition, in the covariantized approach we can find the constraint on the metric when both kinetic and potential terms are present, and this is discussed in section \ref{secV}. This constraint allow us to obtain the effective speed of sound of fluid perturbations, in section \ref{subsecVA}, which can be used to impose conditions on the physically allowed coupling functions.
Then, in section \ref{subsecVB}, we focus our attention on the particular cases leading to adiabatic models. Section \ref{subsecVC} is then devoted to analyzing an open question regarding the possibility of a constant kinetic coupling function, while in section \ref{subsecVD} we consider a particular solution of a family of TDiff theories which yields the same results as GR, and study its stability. Finally, in section \ref{section: Conclusions} we present the main conclusions of the work and discuss further outlook. In Appendix \ref{appendix: Different covariantizations} we discuss a different way of covariantizing our theory, and in Appendix \ref{appendix: Covariantized EMT} we have included some calculations which are not needed to follow the thread of the main discussion, but which the reader might find useful.

As a final note, we remark that our conventions in this work include the usage of units in which $\hbar=c=1$, metric signature $(+, -, -, -)$, and the notation $g = \abs{\det(g_{\mu\nu})}$.

\section{The TDiff approach}\label{section: The TDiff approach}
The breaking of Diff invariance down to TDiff in the matter action was recently studied in general backgrounds in reference \cite{Jaramillo-Garrido}. This work considers a scalar field coupled to gravity via arbitrary functions of the metric determinant, with total action
\begin{equation}\label{eq: model TDiff action}
        S = S_\text{EH} + S_m \,,
\end{equation}
where
\begin{equation}
S_\text{EH} = - \frac{1}{16\pi G} \int d^4x\, \sqrt{g} \, R\,,
\end{equation}
and
\begin{equation}
S_m =  \int d^4x\, \left\{ \frac{f_k(g)}{2} g^{\mu\nu} \partial_\mu \psi \partial_\nu \psi - f_v(g) V(\psi) \right\} \, .
\end{equation}
This action is indeed seen to be invariant only under transverse diffeomorphisms due to the arbitrary functions of (the absolute value of) the metric determinant $f_k(g)$ and $f_v(g)$. In this work we shall assume that $f_k\geq0$ in order to avoid ghost instabilities.

Let us summarize in this section the main results obtained in reference \cite{Jaramillo-Garrido} following the TDiff approach. The equation of motion (EoM) for the scalar field is
\begin{equation}\label{eq: TDiff psi EoM}
    \partial_\mu\big( f_k(g) \partial^\mu \psi \big) + f_v(g) V'(\psi) = 0 \,, 
\end{equation}
where in general a prime denotes derivative with respect to its argument.
On the other hand, the EoM for the gravitational field are the usual Einstein equations
\begin{equation}\label{eq: Einstein eqns}
    G_{\mu\nu} = R_{\mu\nu} - \frac{1}{2}R g_{\mu\nu} = 8\pi G \, T_{\mu\nu} \,,
\end{equation}
where the Energy-Momentum Tensor (EMT) for the scalar field is found from its definition
\begin{equation}\label{eq: def EMT}
     T_{\mu\nu} = \frac{2}{\sqrt{g}} \frac{\delta S_m}{\delta g^{\mu\nu}} \,,
\end{equation}
and reads:
\begin{equation}\label{eq: TDiff EMT}
    \begin{split}
        T_{\mu\nu} &= \frac{2}{\sqrt{g}} \left\{ \frac{1}{2} f_k(g) \partial_\mu \psi \partial_\nu \psi \right. + \\[5pt]
        &\quad + \left.  g \left[ f_v'(g) V(\psi) - \frac{1}{2} f'_k(g) (\partial\psi)^2 \right] g_{\mu\nu} \right\} \,,
    \end{split}
\end{equation}
where we are denoting
\begin{equation}
    (\partial\psi)^2 \equiv \partial_\alpha \psi \partial^\alpha \psi \,.
\end{equation}
Under the assumption of the field derivative $\partial_\mu \psi$ being a timelike vector, it is possible to rewrite the EMT in perfect fluid form. Indeed, defining a unit timelike vector field $u^\mu$ through
\begin{equation}\label{eq: velocity}
     u^\mu = \frac{\partial^\mu \psi}{\sqrt{(\partial\psi)^2}} \equiv \frac{\partial^\mu \psi}{N} \,,
\end{equation}
where we are denoting the normalization as
\begin{equation}
    N \equiv \sqrt{(\partial\psi)^2} \,,
\end{equation}
together with an energy density
\begin{equation}\label{eq: general energy density}
    \rho = \frac{2}{\sqrt{g}} \left\{ \frac{1}{2} f_k (\partial\psi)^2 + g \left[ f_v' V - \frac{1}{2} f'_k (\partial\psi)^2 \right] \right\}
\end{equation}
and a pressure
\begin{equation}\label{eq: general pressure}
    p = -\frac{2g}{\sqrt{g}} \left[ f_v' V - \frac{1}{2} f'_k (\partial\psi)^2  \right] \,,
\end{equation}
one can rewrite the EMT \eqref{eq: TDiff EMT} as
\begin{equation}\label{eq: perfect fluid EMT}
    T_{\mu\nu} = (\rho+p)u_\mu u_\nu -p g_{\mu\nu} \,.
\end{equation}
We shall in this paper also work with the timelike vector assumption whenever we wish to reexpress the analysis as that of a perfect fluid.

One of the main points of study in the TDiff approach is the conservation of the EMT. Indeed, the conservation of this quantity is an automatic consequence of the Noether theorem for a theory with symmetry under diffeomorphisms, but it does not follow trivially when we have less symmetry. Instead, one argues that the conservation of the EMT on the solutions to the equations of motion is a consistency requirement of the theory, since the Einstein equations \eqref{eq: Einstein eqns} still hold and the Einstein tensor is divergenceless:
\begin{equation}\label{consT}
    \nabla_\mu G^{\mu\nu} = 0 \,\implies\, \nabla_\mu T^{\mu\nu} = 0 \,.
\end{equation}
Within the TDiff approach, this consistency condition allows one to obtain a certain (physical) constraint on the metric.

We shall now focus on two limiting cases of interest. These are the potential domination regime and the kinetic domination regime, which we review in the following.

\subsection{Potential domination in the TDiff approach}\label{subsecIIIA}
Everything is rather simple in the potential regime, which amounts to ignoring the kinetic contribution in the action. When we do so, the EoM \eqref{eq: TDiff psi EoM} for $\psi$ becomes $f_v V' = 0$, which (for a nonvanishing coupling function $f_v$) tells us that the field takes on the constant value $\psi=\psi_0$ which is the extremum of the potential: $V(\psi) = V(\psi_0) =$ const. The EMT \eqref{eq: TDiff EMT} simplifies down to
\begin{equation}\label{eq: TDiff EMT potential domination}
    T_{\mu\nu} = 2V f_v' \sqrt{g} \, g_{\mu\nu} \,,
\end{equation}
and its conservation becomes
\begin{equation}\label{eq: TDiff EMT conservation in potential domination}
    2V g^{\mu\nu} \nabla_{\mu}\left( f_v' \sqrt{g} \right) = 2V g^{\mu\nu} \partial_{\mu}\left( f_v' \sqrt{g} \right) = 0\,,
\end{equation}
where in the first term we have pulled (covariantly) constant terms out of the covariant derivative and in the second we have recognized that the product $f_v' \sqrt{g}$ is a TDiff scalar so that we may use a partial derivative. Since relation \eqref{eq: TDiff EMT conservation in potential domination} must be met for all metrics, it follows that
\begin{equation}
    \partial_{\mu}\left( f_v' \sqrt{g} \right) = \frac{1}{\sqrt{g}} \left( \frac{1}{2} f'_v + g f_v'' \right) \partial_\mu g = 0 \,.
\end{equation}
The above relation is satisfied whenever the coupling function takes the form
\begin{eqnarray}\label{eq: condition on fv}
    f_v(g) = A \sqrt{g} + B \,,
\end{eqnarray}
with $A$ and $B$ constants of integration, but if we wish to leave the coupling function arbitrary (which in principle we do), then it must be the case that
\begin{equation}\label{eq: constant determinant}
    \partial_\mu g = 0 \,\implies\, g = \text{constant} \,.
\end{equation}
This is the constraint on the metric which we obtain in the potential domination regime: the determinant must be constant. It is interesting to note that, being the determinant a constant quantity, any given function of the determinant will also assume a constant value, for instance the function $f_v'(g)$. This is relevant because, looking back at the EMT \eqref{eq: TDiff EMT potential domination}, it may be written as
\begin{equation}
    T_{\mu\nu} \equiv \lambda \, g_{\mu\nu} \,,
\end{equation}
with
\begin{equation}
    \lambda \equiv 2V f_v' \sqrt{g} = \text{constant} \,,
\end{equation}
and we have the behaviour of a cosmological constant.

\subsection{Kinetic domination in the TDiff approach}\label{subsecIIIB}
The study of the kinetic domination regime is more involved. The EoM for the field in this regime becomes $\partial_\mu \left(f_k \partial^\mu \psi\right) = 0$, and its solution must satisfy
\begin{equation}\label{eq: TDiff solution to EoM}
    (\partial \psi)^2 = C_\psi(x) \left( \frac{\sqrt{g}}{\delta V f_k}\right)^2 \,,
\end{equation}
with $C_\psi(x)$ a function such that $u^\mu\partial_\mu C_\psi(x) = 0$, and where we find the cross-sectional volume of the congruence $\delta V$, related to the expansion by \cite{Poisson}
\begin{equation}\label{eq: cross-sectional volume}
    \nabla_\mu u^\mu = u^\mu\partial_\mu \ln \delta V \,.
\end{equation}

Regarding EMT conservation, when working with a perfect fluid it is common to project the conservation equations $\nabla_\mu T^{\mu\nu} = 0$ onto directions longitudinal and transverse to the velocity of the fluid. For the former one contracts with $u_\nu$, and for the latter one acts with the orthogonal projector $h\indices{^\mu_\nu} = \delta^\mu_\nu - u^\mu u_\nu$, obtaining respectively:
\begin{subequations}
    \begin{gather}
        \dot{\rho} + (\rho + p) \nabla_\mu u^\mu = 0 \,, \label{subeq: longitudinal conservation}\\[5pt]
        (\rho+p) \dot{u}^\mu - \left( g^{\mu\nu} - u^\mu u^\nu \right) \nabla_\nu p = 0 \,, \label{subeq: transverse conservation}
    \end{gather}
\end{subequations}
where we use the dot notation $\dot{\,\,} \equiv u^\mu \nabla_\mu$. In the kinetic regime, the perfect fluid quantities take on the form
\begin{subequations}
    \begin{align}
        \rho &= \frac{(\partial\psi)^2}{\sqrt{g}} \left( f_k - g f'_k \right) \,, \\[5pt]
        p &= \frac{(\partial\psi)^2}{\sqrt{g}} \, g f'_k \,,
    \end{align}
\end{subequations}
and the equation of state (EoS) parameter reads
\begin{equation}\label{eq: TDiff kinetic EoS}
    w = \frac{p}{\rho} = \frac{g f'_k}{f_k - g f'_k} \equiv \frac{F}{1-F} \,,
\end{equation}
where we are defining
\begin{equation}\label{eq: def F}
    F \equiv \frac{gf'_k}{f_k} \,.
\end{equation}
It is interesting to remark that the EoS parameter \eqref{eq: TDiff kinetic EoS} is a function of only the metric determinant, $w=w(g)$.

Studying the longitudinal projection \eqref{subeq: longitudinal conservation} on the solution to the EoM \eqref{eq: TDiff solution to EoM} yields the following relation:
\begin{equation}\label{eq: TDiff longitudinal constraint}
    (2F-1)\frac{g}{f_k} = C_g(x) \delta V^2 \,,
\end{equation}
with $C_g(x)$ a function which must satisfy
\begin{equation}
    u^\mu\partial_\mu C_g(x) = 0 \,.
\end{equation}
Equation \eqref{eq: TDiff longitudinal constraint} is one of the main results in the TDiff approach, and indeed shows how the study of EMT conservation is not a trivial matter but rather imposes a constraint on the metric.

Probing the transverse projection \eqref{subeq: transverse conservation} for further information ends up revealing that the two functions $C_\psi(x)$ and $C_g(x)$ are inversely proportional, i.e.
\begin{equation}
    C_\psi C_g = \text{constant} \equiv c_\rho \,.
\end{equation}
The reason for naming the constant as $c_\rho$ is because it is possible to find the following nice expression for the energy density in the kinetic regime \cite{Jaramillo-Garrido}:
\begin{equation}\label{eq: TDiff simple rho kinetic}
    \rho = \frac{c_\rho}{(w-1) \sqrt{g}} \,.
\end{equation}
Since the EoS is only a function of the metric determinant, it follows that both the energy density and the pressure are functions of only the metric determinant, and this dependence on a single quantity reveals that we are dealing with an adiabatic fluid. The adiabatic speed of sound is defined through $\delta p = c_a^2 \, \delta\rho$ which, joined with $p=w\rho$, yields
\begin{equation}\label{eq: speed of sound of adiabatic perturbations}
    c_a^2 = w + w' \frac{\rho}{\rho'} \,.
\end{equation}
In the TDiff approach, it takes the form \cite{Jaramillo-Garrido}:
\begin{equation}\label{eq: TDiff speed of sound}
    c_a^2 = -\frac{g f_k (f'_k + 2 g f_k'')}{f_k^2 + (2 g f_k')^2 - gf_k (5 f_k' + 2 g f_k'')} \,.
\end{equation}
In this way we conclude our summary of the general framework resulting from the consideration of a TDiff scalar field. For interesting consequences and phenonomenology of this theory we refer the reader to references \cite{Maroto,Jaramillo-Garrido,Alonso-Lopez}.

\section{Covariantized action}\label{section: Covariantized action}

In this section we note that one can rewrite an action with broken Diff invariance in a Diff invariant way via the introduction of additional fields, similar in spirit to the Stueckelberg procedure in gauge theories \cite{Stueckelberg}. In this work we shall follow Henneaux and Teitelboim \cite{Henneaux} and, in order to preserve locality, introduce the new field in the form of a vector. The way of restoring the Diff symmetry is not unique, however, and other references prefer to introduce a scalar field (see e.g. \cite{Blas:2011ac} and references therein). We refer the reader to Appendix \ref{appendix: Different covariantizations} for a more detailed discussion on this subject.

Having clarified that point, let us now consider a TDiff invariant field theory where each of the terms in the action is of the form
\begin{equation}\label{eq: term in TDiff action}
    S_\text{TDiff}[g_{\mu\nu},\Psi] = \int d^4x\, f(g) \, \mathcal{L}\left( g_{\mu\nu}, \Psi,\partial_\mu\Psi \right) \,,
\end{equation}
with $\mathcal{L}$ a Diff scalar. The general case could be written as the sum of different terms,
\begin{equation}
    S_\text{TDiff}[g_{\mu\nu},\Psi] = \int d^4x\, \sum_i f_i(g) \, \mathcal{L}_i\left( g_{\mu\nu}, \Psi,\partial_\mu\Psi \right) \,,
\end{equation}
but in order not to clutter the treatment we shall consider the simpler expression \eqref{eq: term in TDiff action}, which is precisely a general term in the above summation.

In order to perform the covariantization, we first of all introduce a Diff scalar density $\bar{\mu}$ which transforms as $\sqrt{g}$ under general coordinate transformations. Doing so, the Diff invariant theory given by
\begin{equation}\label{eq: term in Diff action}
    S_\text{Diff}[g_{\mu\nu},\Psi,\bar{\mu}] = \int d^4x\, \sqrt{g} \left[ \frac{\bar{\mu}}{\sqrt{g}}\, f(g/\bar{\mu}^2) \right] \mathcal{L}\left( g_{\mu\nu}, \Psi,\partial_\mu\Psi \right)
\end{equation}
agrees with the TDiff theory \eqref{eq: term in TDiff action} in the coordinate frame in which $\bar{\mu}=1$, which we shall refer to as the ``TDiff frame'' for simplicity. The term in between brackets is a Diff scalar and an arbitrary function of the combination $\bar{\mu}/\sqrt{g}$, which we shall write as
\begin{equation}
    H(\bar{\mu}/\sqrt{g}) \equiv \frac{\bar{\mu}}{\sqrt{g}}\, f(g/\bar{\mu}^2) \,.
\end{equation}
If, for simplicity, we denote the argument by $\bar{\mu}/\sqrt{g} \equiv Y$, then we have that
\begin{equation}\label{eq: defH}
    H(Y) \equiv Y f(Y^{-2}) \,.
\end{equation}
In the TDiff frame $\bar{\mu}=1$ (equivalently, $Y=1/\sqrt{g}$), we would find
\begin{equation}\label{eq: gauge}
    H(Y) \bigg|_{\bar{\mu}=1} = \frac{f(g)}{\sqrt{g}}
\end{equation}
The question now is how the newly introduced scalar density $\bar{\mu}$ is related to the new field (equivalently, how the combination $Y$ is related to the new field). As we mentioned there are different possibilities, but we shall choose a vector field $T^\mu$ as our addition to the theory. A scalar density $\bar{\mu}$ which transforms as $\sqrt{g}$ may be built from a vector field through the simple combination \cite{Maroto,Henneaux}
\begin{equation}
    \bar{\mu} = \partial_\mu \left(\sqrt{g} \, T^\mu\right) \,.
\end{equation}
In this way, the variable $Y$ turns out to be related to $T^\mu$ through
\begin{equation}
    Y=\frac{\bar{\mu}}{\sqrt{g}} = \nabla_\mu T^\mu \,,
\end{equation}
and so we see that the newly introduced vector field $T^\mu$ enters the theory via its divergence. It is interesting to remark that in the general Stueckelberg procedure the idea is to take the gauge functions and promote them to fields. Since 4-dimensional diffeomorphisms are generated by 4 functions, it is natural to introduce a vector field $T^\mu$. This being said, however, going from TDiff back to Diff actually only involves removing one condition (recall $J=1$, or $\partial_\mu \xi^\mu = 0$), and thus it makes sense that the newly introduced vector field ends up appearing through the scalar combination $Y=\nabla_\mu T^\mu$.

Our covariantized action \eqref{eq: term in Diff action} would then take the form
\begin{equation}
    S_\text{Diff}[g_{\mu\nu},\Psi,T^\mu] = \int d^4x\, \sqrt{g} \, H(Y)\, \mathcal{L}\left( g_{\mu\nu}, \Psi,\partial_\mu\Psi \right) \,,
\end{equation}
where we bear in mind that $Y=\nabla_\mu T^\mu$. We note that in the case $H(Y) = \text{const.}$ the dependence on the new field would be lost because it would correspond to the original theory being Diff invariant already (i.e. $f(g) \propto \sqrt g$). In any case, since
\begin{equation}
    S_\text{Diff}[g_{\mu\nu},\Psi,T^\mu] \bigg|_{\bar{\mu}=1} = S_\text{TDiff}[g_{\mu\nu},\Psi] \,,
\end{equation}
we see how one may alternatively work within the TDiff approach or the covariantized approach since they are equivalent, as it is explicitly shown when choosing the gauge $\bar{\mu}=1$ (equivalently, $Y=1/\sqrt{g}$) in the Diff invariant action. Thus, the invariance under full diffeomorphisms has been restored due to the introduction of a Diff vector field $T^\mu$.

\section{The covariantized approach}\label{section: The covariantized approach}
Let us take the Diff invariant action \cite{Maroto}
\begin{equation}\label{eq: covariantized model action}
    S_\text{Diff}[g_{\mu\nu},\psi,T^\mu] = S_\text{EH} + \int d^4x\, \sqrt{g} \left[ H_k(Y) X - H_v(Y) V  \right]
\end{equation}
where, for simplicity, we are denoting the kinetic term by
\begin{equation}
    X \equiv \frac{1}{2} g^{\mu\nu} \partial_\mu \psi \partial_\nu\psi = \frac{1}{2} (\partial\psi)^2 \,.
\end{equation}
If we set $\bar{\mu}=1$ and recognize the functions $f_k(g)$ and $f_v(g)$, we immediately realize that it has become precisely action \eqref{eq: model TDiff action} for a TDiff scalar field. Let us keep working now in the covariantized approach, and see what we obtain.
Variations of action \eqref{eq: covariantized model action} with respect to the scalar field $\psi$ yield the following EoM
\begin{equation}\label{eq: covar EoM psi}
  \nabla_\nu \left[ H_k(Y) g^{\mu\nu} \nabla_\mu\psi \right] + H_v(Y) V'(\psi) = 0\, .
\end{equation}
Recalling that the covariant divergence of a vector $V^\alpha$ may be expressed as
\begin{equation}
    \nabla_\alpha V^\alpha = \frac{1}{\sqrt{g}}\partial_\alpha\left(\sqrt{g} \, V^\alpha \right) \,,
\end{equation}
and using equation \eqref{eq: gauge}, one can see that in the $\bar{\mu} = 1$ gauge this equation reduces to
\begin{equation}
    \frac{1}{\sqrt{g}}\partial_\nu [f_k(g)\partial^\mu\psi]+\frac{f_v(g)}{\sqrt{g}}V'(\psi)=0\, ,
\end{equation}
which is equivalent to the EoM \eqref{eq: TDiff psi EoM} we found for $\psi$ in the TDiff approach.

On the other hand, variations of action \eqref{eq: covariantized model action} with respect to the vector field that has restored the Diff invariance, that is $T^\mu$, provide the following EoM:
\begin{equation}\label{eq: EoM Tmu}
    \partial_\nu \left[ H_k'(Y) X - H_v'(Y) V \right] = 0 \,.
\end{equation}
Hence, there is a conserved quantity as a result of the derivative dependence of the field $T^\mu$ on action \eqref{eq: covariantized model action}. One could think that these equations are new with respect to the TDiff formalism. However, it should be noted that in said formalism one has the equations coming from the conservation of the EMT that is no longer trivially satisfied in that framework, which are equations \eqref{consT}. So, as we will show, these equations must be equivalent.

Finally, variations with respect to the metric tensor $g_{\mu\nu}$ provide the Einstein equations
\begin{equation}
    G_{\mu\nu} = 8\pi G \, T_{\mu\nu} \,,
\end{equation}
where the total EMT is found using definition \eqref{eq: def EMT} and takes the form
\begin{equation}\label{eq: covariantized EMT}
    \begin{split}
        T_{\mu\nu} &= H_k(Y) \partial_\mu \psi \partial_\nu \psi - \left[ H_k(Y) X - H_v(Y) V \right] g_{\mu\nu} \\[5pt]
        &\qquad + Y \left[ H_k'(Y) X - H_v'(Y) V \right] g_{\mu\nu} 
    \end{split}
\end{equation}
(see Appendix \ref{appendix: Covariantized EMT} for the calculation of this quantity). It should be noted that the conservation of the EMT \eqref{eq: covariantized EMT} has to be trivially satisfied when considering the EoM of the fields \eqref{eq: covar EoM psi} and \eqref{eq: EoM Tmu}, due to the invariance of action \eqref{eq: covariantized model action} under general diffeomorphisms. Indeed, it can be explicitly checked that this is the case, and we refer the reader to Appendix \ref{appendix: Covariantized EMT} for the calculation.

On the other hand, noting the definition of the $H(Y)$ functions in equation \eqref{eq: defH}, one can obtain the relation
\begin{equation}
        H'(Y)=
    f\left(Y^{-2}\right) - 2 Y^{-2} \frac{df}{d\left(Y^{-2}\right)} \,,
\end{equation}
so that
\begin{equation}\label{eq: H' in TDiff frame}
    H'(Y) \bigg|_{\bar{\mu}=1} = f(g) - 2g f'(g) \,.
\end{equation}
Taking into account this result, together with equation \eqref{eq: gauge}, it is immediate to verify that the covariantized EMT \eqref{eq: covariantized EMT} reduces to the TDiff EMT \eqref{eq: TDiff EMT} in the $\bar{\mu}=1$ frame.

This covariantized formalism has been applied to the FLRW model in reference \cite{Maroto}, recovering the information provided by the constraint on the metric in the TDiff case. In the next section we will show that this is possible in general.

\subsection{Potential domination in the covariantized approach}\label{subsecIVA}
We begin with the potential domination regime. Neglecting the kinetic parts, the EoM \eqref{eq: covar EoM psi} for $\psi$ becomes
\begin{equation}
    H_v(Y) V'(\psi) = 0 \,,
\end{equation}
and this means that either $H_v = 0$ (which is trivial) or that the field takes on the constant value $\psi(x) =  \psi_0$ such that the potential reaches an extremum $V(\psi) = V(\psi_0) =$ const. (exactly as we concluded in the TDiff approach). On the other hand, the EoM \eqref{eq: EoM Tmu} for $T^\mu$ in the potential regime become
\begin{equation}
    \partial_\mu \left[ H_v'(Y) V \right] = 0 \,\Rightarrow\, \partial_\mu H_v'(Y) = H_v''(Y) \, \partial_\mu Y = 0 \,,
\end{equation}
where we have already used that $V=$ constant to pull it out of the derivative. We now have a couple of options in order for this equation to be satisfied, namely that $H_v''(Y) = 0$ or that $\partial_\mu Y = 0$. The first of these implies
\begin{equation}
    H_v''(Y) = 0 \,\implies\, H_v(Y) = BY + A \,,
\end{equation}
which in the $\bar{\mu}=1$ frame, that is $H_v\rightarrow f_v/\sqrt{g}$ and $Y\rightarrow 1/\sqrt{g}$, reduces to the condition \eqref{eq: condition on fv} on the coupling function $f_v$ we already found. The second option gives
\begin{equation}
    \partial_\mu Y = 0 \,\implies\, Y = \text{constant} \,,
\end{equation}
which in the $\bar{\mu}=1$ frame reduces to $g=$ constant, and this is precisely the condition \eqref{eq: constant determinant} on the metric determinant we previously obtained.

Moreover, the EMT \eqref{eq: covariantized EMT} in the potential regime takes the form
\begin{equation}
    T_{\mu\nu} = V\left(H_v - Y H_v'\right) g_{\mu\nu} \equiv \lambda \, g_{\mu\nu}\,,
\end{equation}
and if we use the above EoM we see that it is of a cosmological constant-type once again, since
\begin{equation}
    \lambda \equiv V\left(H_v - Y H_v'\right) = \text{constant} \,.
\end{equation}
Thus, in the potential limit we have recovered our previous results, as it should be expected.

\subsection{Kinetic domination in the covariantized approach}\label{subsecIVB}
In the kinetic domination regime we neglect the potential terms. The EoM \eqref{eq: EoM Tmu} for $T^\mu$ become
\begin{equation}\label{eq: covar EoM Tmu kinetic}
    \partial_\mu \left[ H_k'(Y) X \right] = \frac{1}{2} \partial_\mu \left[ H_k'(Y) (\partial\psi)^2 \right] = 0 \,.
\end{equation}
It follows from the above that
\begin{equation}\label{eq: solution to covar EoM Tmu kinetic}
    H_k'(Y) (\partial\psi)^2 = \text{constant} \equiv -c_\rho \,,
\end{equation}
where we have named the arbitrary constant as $-c_\rho$ for future convenience. On the other hand, the EoM \eqref{eq: covar EoM psi} for $\psi$ in the kinetic domination regime becomes
\begin{equation}\label{eq: covar EoM psi kinetic}
    \nabla_\mu \left[ H_k(Y) \partial^\mu\psi \right] =  \nabla_\mu \left[ H_k(Y) N u^\mu \right] = 0\, ,
\end{equation}
where we have recalled the velocity \eqref{eq: velocity}. If we expand the equation and divide through by $H_k$ (which we take to be nonzero), then we obtain
\begin{equation}
    N u^\mu \nabla_\mu\ln H_k + u^\mu \nabla_\mu N + N \nabla_\mu u^\mu = 0 \,.
\end{equation}
Recalling equation \eqref{eq: cross-sectional volume} for the relation between the expansion $\nabla_\mu u^\mu$ and the cross-sectional volume $\delta V$, and using the fact that the covariant derivatives may be changed by partial derivatives when acting on scalar functions, it follows from the above that
\begin{equation}
    \begin{split}
        N u^\mu \partial_\mu \left( \ln H_k \right.&+ \left.\ln N + \ln \delta V \right) =\\[5pt]
        &= N u^\mu \partial_\mu \ln\left(H_k N \delta V\right) = 0 \,.
    \end{split}
\end{equation}
From this expression, 
we finally conclude that
\begin{equation}\label{eq: covar EoM psi kinetic solutions}
    (\partial\psi)^2 = \frac{C_\psi(x)}{(H_k \delta V)^2} \,,
\end{equation}
with $C_\psi(x)$ a function such that $u^\mu \partial_\mu C_\psi(x) = 0$. Going now to the $\bar{\mu}=1$ frame and using equation \eqref{eq: gauge} for $H_k|_{\bar\mu=1}$, we immediately recover expression \eqref{eq: TDiff solution to EoM} for the solution to the EoM in the TDiff approach.
 
On the other hand, substituting equation \eqref{eq: covar EoM psi kinetic solutions} into \eqref{eq: solution to covar EoM Tmu kinetic}, it follows that
\begin{equation}
     -c_\rho = H_k' (\partial\psi)^2 = H_k' \frac{C_\psi(x)}{(H_k \delta V)^2} \, ,
\end{equation}
which leads to
\begin{equation}
     -\frac{H_k'}{H_k^2} = \frac{c_\rho}{C_\psi(x)}\, \delta V^2 \,.
\end{equation}
If we now define the function
\begin{equation}
    C_g(x) \equiv \frac{c_\rho}{C_\psi(x)} \,,
\end{equation}
which satisfies $u^\mu \partial_\mu C_g(x) = 0$, we may write the above equation as
\begin{equation}
    -\frac{H_k'}{H_k^2} = C_g(x) \delta V^2 \,.
\end{equation}
It is now immediate to see that we recover the constraint \eqref{eq: TDiff longitudinal constraint} when going to the frame $\bar{\mu}=1$, since
\begin{equation}
    \left. -\frac{H_k'}{H_k^2}  \right|_{\bar{\mu}=1} = -\frac{f_k - 2g f_k'}{f_k^2\,/\,g} \, ,
    \end{equation}
and this implies that
 \begin{equation}   
    (2F-1)\frac{g}{f_k}= C_g(x) \delta V^2\,,
\end{equation}
where we have recalled the definition of the function $F$ in equation \eqref{eq: def F}. It is worth noting that obtaining this result in the TDiff approach requires a longer calculation \cite{Jaramillo-Garrido}, while in the covariantized approach we have found it rather directly.

On another note, in the kinetic domination regime the EMT \eqref{eq: covariantized EMT} takes the form
\begin{equation}
    T_{\mu\nu} = H_k \partial_\mu \psi \partial_\nu \psi - \left[ H_k - Y H_k' \right] X g_{\mu\nu} \,,
\end{equation}
which, under the assumption of the field derivative $\partial_\mu\psi$ being a timelike vector, is seen to be equivalent to that of a perfect fluid, using the same definition \eqref{eq: velocity} for the velocity, and with energy density
\begin{equation}\label{eq: covar rho kinetic}
    \rho = -\frac{c_\rho}{2} \left( \frac{H_k}{H_k'} + Y \right)
\end{equation}
and pressure
\begin{equation}
    p = -\frac{c_\rho}{2} \left( \frac{H_k}{H_k'} - Y \right) \,,
\end{equation}
where we have made use of the solution \eqref{eq: solution to covar EoM Tmu kinetic} to simplify $H_k' X = -c_\rho/2$. Note that we have obtained that the energy density and pressure are functions of only $Y$, and this dependence on a single variable means that possible perturbations of the fluid will be adiabatic.

The EoS parameter now reads
\begin{equation}\label{eq: wcov}
    w = \frac{p}{\rho} = \frac{\frac{H_k}{H_k'} - Y}{\frac{H_k}{H_k'} + Y} \,,
\end{equation}
and, as expected, is seen to be a function of $Y$ only.
Note also that in the $\bar{\mu}=1$ frame we have
\begin{equation}\label{eq: recover EoS}
    \left. w \right|_{\bar{\mu}=1}
     = \frac{  \frac{f_k/\sqrt{g}}{f_k-2gf'_k} -\frac{1}{\sqrt{g}} }{  \frac{f_k/\sqrt{g}}{f_k-2gf'_k} +\frac{1}{\sqrt{g}}}=\frac{gf'_k}{f_k-gf'_k}\,,
\end{equation}
which coincides with expression \eqref{eq: TDiff kinetic EoS} for the EoS parameter previously found in the TDiff formalism. Consider now the following quantity:
\begin{equation}
    w-1 = \frac{-2 Y}{\frac{H_k}{H_k'} + Y} = \frac{Y c_\rho}{\rho} \,,
\end{equation}
where in the second equality we have recalled expression \eqref{eq: covar rho kinetic} for the energy density. Rearranging, we obtain the following simple expression for the energy density:
\begin{equation}
    \rho = \frac{Y c_\rho}{w-1} \,.
\end{equation}
After evaluation in the frame $\bar{\mu}=1$, i.e. substituting $Y\rightarrow 1/\sqrt{g}$ and recalling from equation \eqref{eq: recover EoS} that the EoS parameter is recovered, it finally yields the simple expression \eqref{eq: TDiff simple rho kinetic} for the energy density that was obtained in the TDiff approach.

Finally, the speed of sound of the adiabatic perturbations follows from the definition \eqref{eq: speed of sound of adiabatic perturbations}, which yields
\begin{equation}\label{eq: cacov v1}
    c_a^2 = \frac{H_k \, H_k''}{H_k \, H_k'' - 2 \left(H_k'\right)^{2}} \,,
\end{equation}
or, finding a common factor,
\begin{equation}\label{eq: cacov}
    c_a^2 = \frac{1}{1 - 2 \frac{\left(H_k'\right)^{2}}{H_k \, H_k''}} \,.
\end{equation}
Note that in obtaining this second expression we have divided by $H_k''$; so, it will not be valid, for example, for the interesting case of TDiff dark matter that will be commented in the next section. From this equation it is easy to note that, in order have a stable adiabatic fluid (i.e. a non-negative $c_a^2$), one needs the coupling function to satisfy\footnote{The case in which $H_k''=0$ implies $c_a^2 = 0$, which trivially satisfies both stability and subluminality requirements. Hence, what we present here implicitly assumes that $H_k''\neq 0$.}
\begin{equation}\label{eq: stability in covar kinetic}
    \frac{\left(H_k'\right)^{2}}{H_k \, H_k''} < \frac{1}{2} \,.
\end{equation}
Moreover, in order to also avoid the propagation of superluminal perturbations (i.e. to have $c_a^2 \leq 1$), it must be the case that
\begin{equation}\label{eq: subluminal in covar kinetic}
        H_k''< 0 \,,
\end{equation}
where we have taken into account that $H_k>0$ (so that the kinetic term is positive and there are no ghosts in the theory).

On the other hand, it is possible once again to verify that equation \eqref{eq: cacov}, when evaluated in the $\bar{\mu}=1$ frame, reduces to the previously obtained adiabatic speed of sound in equation \eqref{eq: TDiff speed of sound}. To this end, we must simply recall equations \eqref{eq: gauge} and \eqref{eq: H' in TDiff frame} for $ H|_{\bar{\mu}=1}$ and $H'|_{\bar{\mu}=1}$, respectively, and also use that
\begin{equation}
    H''(Y) = 2 Y^{-3} \left[ \frac{df}{d\left(Y^{-2}\right)} + 2 Y^{-2} \frac{d^2f}{d^2\left(Y^{-2}\right)} \right] \,,
\end{equation}
which yields
\begin{equation}
    H''(Y) \bigg|_{\bar{\mu}=1} = 2 g^{3/2} \big[ f'(g) + 2g f''(g) \big] \,.
\end{equation}
After careful substitution, we finally obtain 
\begin{equation}\label{eq: TDiff speed of sound again}
    \left. c_a^2 \right|_{\bar{\mu}=1} = -\frac{g f_k (f'_k + 2 g f_k'')}{f_k^2 + (2 g f_k')^2 - gf_k (5 f_k' + 2 g f_k'')} \,,
\end{equation}
which indeed recovers the TDiff adiabatic speed of sound in equation \eqref{eq: TDiff speed of sound}.

Furthermore, evaluated in the $\bar{\mu}=1$ frame the requirement for stability \eqref{eq: stability in covar kinetic} is translated to
\begin{equation}\label{eq: stability in TDiff kinetic}
    \frac{(f_k - 2g f_k')^2}{g f_k (f_k' + 2g f_k'')} < 1 \,,
\end{equation}
while the requirement for subluminal propagation \eqref{eq: subluminal in covar kinetic} becomes
\begin{equation}
    f_k' + 2g f_k'' < 0 \,.
\end{equation}
These conditions on the kinetic coupling function $f_k$ can also be seen to follow directly from the TDiff expression for the adiabatic speed of sound \eqref{eq: TDiff speed of sound again}, as should be expected. In order to see precisely how, it comes in handy to rewrite said expression in a cleaner manner. Inspired by the form of the quantities appearing in equation \eqref{eq: stability in TDiff kinetic}, it is possible to see how the denominator of \eqref{eq: TDiff speed of sound again} can be written as
\begin{equation}
    \begin{split}
    f_k^2 &+ (2 g f_k')^2 - gf_k (5 f_k' + 2 g f_k'') = \\[5pt]
    &\quad = (f_k - 2g f_k')^2 - g f_k (f_k' + 2 g f_k'') \,.
    \end{split}
\end{equation}
Knowing this, it turns out that we can rewrite the TDiff adiabatic speed of sound \eqref{eq: TDiff speed of sound again} in the rather clean form
\begin{equation}
    \left. c_a^2 \right|_{\bar{\mu}=1} = \frac{1}{1 - \frac{(f_k - 2g f_k')^2}{g f_k (f_k' + 2g f_k'')}} \,,
\end{equation}
from which, indeed, the same conditions can easily be seen to follow. Nevertheless, it is worth stressing that knowing the results in the covariantized approach and translating them into the TDiff approach is precisely what helped us in finding an appropriate and simple rewrite for the TDiff adiabatic speed of sound. Indeed, directly working from the convoluted expression \eqref{eq: TDiff speed of sound again} would have implied a much greater effort in finding the conditions for stability and subluminality, and in fact this work was not done in reference \cite{Jaramillo-Garrido}.

As a general comment before concluding, we remark that the above conditions on either $H_k$ or $f_k$ (which are the requirements for stable perturbations and subluminal propagation) serve as a way of selecting classes of possible coupling functions (and hence particular models) which are physically meaningful. Of course, there is in principle no reason to restrict ourselves to only those two requirements, and one could also consider the study of the energy conditions (as was done in reference \cite{Jaramillo-Garrido} in the TDiff framework) as a way of further restricting physically viable theories.

\subsection{Simple models in the kinetic regime}\label{subsecIVC}
Regarding some particular models of interest in the kinetic regime, note that one can obtain a constant equation of state parameter for the fluid considering a power-law kinetic function. Indeed, taking into account equation \eqref{eq: wcov} one has
\begin{equation}
    H_k(Y)=CY^\beta \,\implies\, w=\frac{1-\beta}{1+\beta} \,.
\end{equation}
As it was already discussed in reference \cite{Jaramillo-Garrido} following the TDiff approach, in this case we have that the propagation speed of the adiabatic field perturbations is
\begin{equation}
    c_a^2=\frac{1-\beta}{1+\beta}=w \,,
\end{equation}
which has been obtained taking into account equation \eqref{eq: cacov}. This is because a power-law kinetic function in the covariantized approach implies a power-law kinetic coupling in the TDiff approach but with a different value of the exponent (note equation \eqref{eq: defH}). It should be  noted that the dark matter model commented in reference \cite{Jaramillo-Garrido} in this framework corresponds to $H_k=CY$.

However, the models with a constant equation of state parameter are the only family that will have a similar functional expression in both approaches.
For example, an exponential kinetic function in the covariantized approach is not equivalent to an exponential kinetic coupling in the TDiff approach according to equation \eqref{eq: defH}. In this case, equation \eqref{eq: wcov} leads to the equation of state parameter
\begin{equation}
   H_k(Y)=Ce^{\beta Y} \,\implies\,   w=\frac{1-\beta Y}{1+\beta Y} \, ,
\end{equation}
which could interpolate between the behaviour of stiff matter and that of a cosmological constant.
Nevertheless, taking into account equation \eqref{eq: cacov}, one obtains 
\begin{equation}
    c_a^2 =-1 \,.
\end{equation}
So, the particular models with an exponential kinetic function in the covariantized approach are unstable, whereas models with an exponential kinetic coupling in the TDiff approach can have interesting phenomenology \cite{Alonso-Lopez}. Finally, thus, it should be emphasized that models which could be natural to consider in one approach (due to a simple form of the relevant function) could appear unnatural in the other approach (in which the relevant function could be more complicated).

\section{General models}\label{secV}
Up to this point we have focused on the two limiting regimes and recovered our previously known results, explicitly confirming the equivalence between the two approaches. However, the covariantized treatment can go further: indeed, it yields in a straightforward manner a very important and previously unknown result, namely, the general constraint on the metric when both kinetic and potential terms are present. Let us now discuss this point.

In the TDiff approach of reference \cite{Jaramillo-Garrido} the analysis is restricted to the two limiting regimes of kinetic and potential domination for simplicity, because the study of the conservation of the TDiff EMT \eqref{eq: TDiff EMT} requires an incredible amount of effort. In this way, the constraints are obtained in each of the two regimes, but the general situation with both kinetic and potential terms is not discussed. However, in the covariantized approach, the EoM of the new vector field \eqref{eq: EoM Tmu} are very simple to study, much more so than the conservation of the TDiff EMT; moreover, as we discussed, they encode the same information. Solving those EoM, then, we will obtain the metric constraint in the most general situation.

Now, the solution to the EoM \eqref{eq: EoM Tmu} of the new field $T^\mu$ reads, quite simply,
\begin{equation}
    H_k' X - H_v' V = \text{const.}
\end{equation}
The value of the constant is in principle ``arbitrary'' (speaking more precisely, it is fixed by the initial conditions), but since we should recover the results we already know from the limiting regimes, we will write it as
\begin{equation}\label{eq: EoM Tmu solutions}
    H_k' X - H_v' V = -\frac{c_\rho}{2} \,,
\end{equation}
so that the kinetic limit is indeed immediately verified (we remark that the aforementioned ``arbitrariness'' is not lost, it simply resides in the constant $c_\rho$ which is fixed by the initial conditions\footnote{We can see from equation \eqref{eq: EoM Tmu solutions} how giving initial conditions for the fields and their derivatives fixes the constant $c_\rho$. Note also that an initial condition for $Y$ really is, in the TDiff frame, just an initial condition for $g$; this makes more sense in the context of equation \eqref{eq: general metric constraint}.}). If we now evaluate the above expression in the $\bar{\mu}=1$ frame, taking into account equation \eqref{eq: H' in TDiff frame} for $\left. H' \right|_{\bar{\mu} = 1}$, then we obtain the following:
\begin{equation}\label{eq: general metric constraint}
    (f_k - 2gf_k') X - (f_v - 2gf_v') V = -\frac{c_\rho}{2} \,.
\end{equation}
This novel result is the general expression for the constraint on the metric whenever we have both a kinetic and a potential term present, something which was not previously studied. It is easy to verify that it gives the correct results in the two limiting regimes. Of course, it is a formal expression, meaning that one would need to solve the EoM for the scalar field $\psi$ in order to extract actual information on the metric constraint. Nevertheless, the main point we wish to highlight is the great utility of the covariantized approach, as it has yielded a general and previously unknown result.

On another note, the simple solution \eqref{eq: EoM Tmu solutions} to the EoM for $T^\mu$ also helps in simplifying the covariantized EMT \eqref{eq: covariantized EMT}, in particular its last term, so that in the end we may write it as
\begin{equation}
    T_{\mu\nu} = H_k \partial_\mu \psi \partial_\nu \psi - \left( H_k X - H_v V + \frac{c_\rho}{2} Y \right) g_{\mu\nu} \,.
\end{equation}
Assuming a timelike derivative $\partial_\mu \psi$, we may again express it in perfect fluid form using the usual definition \eqref{eq: velocity} for the velocity and defining the energy density and pressure, respectively, as
\begin{subequations}\label{eq: covar energy density and pressure}
    \begin{align}
        \rho &= H_k X + H_v V - \frac{c_\rho}{2} Y \,, \label{subeq: rho covar}\\[5pt]
        p &= H_k X - H_v V + \frac{c_\rho}{2} Y \,, \label{subeq: p covar}
    \end{align}
\end{subequations}
while the EoS parameter reads
\begin{equation}\label{eq: covar EoS general case}
    w = \frac{H_k X - H_v V + Y c_\rho/2}{H_k X + H_v V - Y c_\rho/2} \,.
\end{equation}
Let us remark that, as these expressions stand, we cannot in general ensure the adiabaticity of our perfect fluid (see the following sections for further comments on this).

Finally, let us briefly discuss the GR limit of our theory, which essentially means $f_k = f_v \equiv f = \sqrt{g}$ in the TDiff approach or, equivalently, $H_k = H_v \equiv H = 1$ in the covariantized approach. For simplicity, we shall only discuss it within the covariantized approach in order to keep it brief, since they are equivalent. Approaching the GR limit is done by considering small variations around the GR solution, which we shall write as
\begin{equation}
    H(Y) = 1 + \epsilon \, h(Y) \,,
\end{equation}
with $\epsilon$ a small parameter for power-counting and $h(Y)$ an arbitrary function, and then taking the limit $\epsilon \to 0$. Now, equation \eqref{eq: EoM Tmu solutions} tells us that
\begin{equation}
    \epsilon (X - V) h'  = -\frac{c_\rho}{2} \,,
\end{equation}
so that the arbitrary constant is $c_\rho =\order{\epsilon}$. Looking at the energy density and pressure in \eqref{eq: covar energy density and pressure}, this means that when approaching the GR limit these quantities behave as
\begin{subequations}
    \begin{align}
        \rho &= X + V + \order{\epsilon} \,, \\[5pt]
        p &= X - V + \order{\epsilon} \,,
    \end{align}
\end{subequations}
and the EoS parameter as
\begin{equation}
    w \underset{\epsilon \to 0}{=} \frac{X - V }{X + V  } \,.
\end{equation}
Thus, we indeed recover for all of these quantities the same expression that we would have for a canonical scalar field in GR. In the subsequent sections, we shall also verify the GR limit at different points.

\subsection{Effective speed of sound}\label{subsecVA}
We mentioned previously how we could not in general ensure the adiabaticity of the fluid. In this section we will see that the most general situation is precisely a non-adiabatic fluid. A comment is needed before proceeding, however: we shall assume from this point onwards that the kinetic coupling function $H_k$ is not constant, so that $H_k'\neq 0$. The particular case of a constant coupling function requires a separate study which the reader may find in the final subsection.

Having cleared that up, let us now begin by considering the solutions \eqref{eq: EoM Tmu solutions} to the EoM of $T^\mu$. Solving for the kinetic term and showing every dependence, we obtain
\begin{equation}\label{eq: X(Y,psi)}
    X = \frac{H_v'(Y) V(\psi) - c_\rho/2}{H_k'(Y)} = X(Y,\psi) \,.
\end{equation}
Explicitly substituting this expression in equations \eqref{eq: covar energy density and pressure} we obtain that the energy density and pressure satisfy
\begin{subequations}\label{eq: covar energy density and pressure after substituting X(Y,psi)}
    \begin{align}
        \rho &= \frac{H_k}{H_k'} \left( H_v' V - \frac{c_\rho}{2} \right) + H_v V - \frac{c_\rho}{2} Y = \rho(Y,\psi) \,,\\[5pt]
        p &= \frac{H_k}{H_k'} \left( H_v' V - \frac{c_\rho}{2} \right) - H_v V + \frac{c_\rho}{2} Y = p(Y,\psi)\,,
    \end{align}
\end{subequations}
i.e. they are functions of two variables and so our fluid is not adiabatic. Now, as functions of two variables, the perturbations will be written
\begin{subequations}
    \begin{align}
        \delta\rho &= \left. \frac{\partial \rho}{\partial Y} \right|_\psi \delta Y + \left. \frac{\partial \rho}{\partial \psi} \right|_Y \delta \psi \,,\\[5pt]
        \delta p &= \left. \frac{\partial p}{\partial Y} \right|_\psi \delta Y + \left. \frac{\partial p}{\partial \psi} \right|_Y \delta \psi\,,
    \end{align}
\end{subequations}
and we may join these equations to write\footnote{Strictly speaking, we should take into account the perturbations in the value of the constant $c_\rho$. However, since it is a constant, it will contribute only to the zero mode; as such, it will not affect the subsequent discussion regarding the propagation of the perturbations.}
\begin{equation}\label{eq: general delta p}
    \delta p = c_s^2 \, \delta\rho + \alpha \, \delta\psi \,,
\end{equation}
where we denote
\begin{subequations}
    \begin{align}
        c_s^2 &\equiv \left. \frac{\partial p / \partial Y}{\partial \rho / \partial Y} \right|_\psi \,, \label{subeq: def cs2}\\[5pt]
        \alpha &\equiv (-c_s^2) \left. \frac{\partial \rho}{\partial \psi} \right|_Y + \left. \frac{\partial p}{\partial \psi} \right|_Y \,.
    \end{align}
\end{subequations}
Note that $c_s^2$ will be the effective speed of sound of cosmological perturbations. Indeed, in the reference frame comoving with the fluid (sometimes called the “rest” frame) we would find $\delta\psi = 0$ and also $\delta p_\text{rest} = c_s^2 \delta\rho_\text{rest}$ (see references \cite{Hu:1998kj,Gordon:2004ez} for a discussion). On another note, in situations in which $\alpha = 0$, then $c_s^2$ would play the role of the adiabatic speed of sound (indeed, recalling that for adiabatic perturbations $\delta p = c_a^2 \delta\rho$, we would have that $c_a^2 = c_s^2$).

Having mentioned those physical interpretations, let us now compute $c_s^2$ from its definition in equation \eqref{subeq: def cs2}. To this end, we shall proceed bit by bit, starting with the numerator. Differentiating the pressure as expressed in equation \eqref{eq: covar energy density and pressure}, we get
\begin{equation}\label{eq: partial p}
    \begin{split}
        \left. \frac{\partial p}{\partial Y} \right|_\psi &= H_k' X + H_k \left. \frac{\partial X}{\partial Y} \right|_\psi - H_v' V + \frac{c_\rho}{2} =\\[5pt]
        &= H_k \left. \frac{\partial X}{\partial Y} \right|_\psi \,,
    \end{split}
\end{equation}
where in the second equality we have used the solutions \eqref{eq: EoM Tmu solutions} to the EoM of the vector field $T^\mu$. Now, differentiating equation \eqref{eq: X(Y,psi)} it follows that
\begin{equation}
    \left. \frac{\partial X}{\partial Y} \right|_\psi = - \frac{1}{\left(H_k'\right)^2} \left[ V \left(H_k'' H_v' - H_k' H_v'' \right) - \frac{c_\rho}{2} H_k'' \right]
\end{equation}
and so, finally, we have that the numerator of $c_s^2$ in equation \eqref{subeq: def cs2} takes the form
\begin{equation}\label{eq: partial p resp Y}
    \left. \frac{\partial p}{\partial Y} \right|_\psi = - \frac{H_k}{\left(H_k'\right)^2} \left[ V \left(H_k'' H_v' - H_k' H_v'' \right) - \frac{c_\rho}{2} H_k'' \right] \,.
\end{equation}
Consider now the denominator of $c_s^2$ in equation \eqref{subeq: def cs2}. Differentiating the energy density \eqref{eq: covar energy density and pressure}, we obtain
\begin{equation}
    \begin{split}
        \left. \frac{\partial \rho}{\partial Y} \right|_\psi &= H_k' X + H_k \left. \frac{\partial X}{\partial Y} \right|_\psi + H_v' V - \frac{c_\rho}{2} =\\[5pt]
        &= 2 \left( H_v' V - \frac{c_\rho}{2} \right) + \left. \frac{\partial p}{\partial Y} \right|_\psi \,,
    \end{split}
\end{equation}
where in the second equality we have used the solutions \eqref{eq: EoM Tmu solutions} to rewrite $H_k' X$, and also recalled expression \eqref{eq: partial p}. Now that we have both the numerator and the denominator of \eqref{subeq: def cs2}, we can compute the quantity $c_s^2$ to be
\begin{equation}\label{eq: cs2 in terms of A and B original}
    c_s^2 = \frac{A(Y,\psi)}{A(Y,\psi) - B(Y,\psi)} \,,
\end{equation}
where we are denoting
\begin{subequations}\label{eq: A(Y,psi) and B(Y,psi)}
    \begin{align}
        A(Y,\psi) &\equiv H_k \left[ V \left( H_k'' H_v' - H_k' H_v'' \right) - \frac{c_\rho}{2} H_k'' \right] \,,\\[5pt]
        B(Y,\psi) &\equiv 2(H_k')^2 \left( H_v' V - \frac{c_\rho}{2} \right) = 2(H_k')^3 X(Y,\psi)\,.
    \end{align}
\end{subequations}
(Note that we have used expression \eqref{eq: X(Y,psi)} in the final equality in order to write $B(Y,\psi)$ in a more compact manner). We may rewrite the above expression for $c_s^2$ in a cleaner form in cases where $A(Y,\psi) \neq 0$ by using it as a common factor, obtaining
\begin{equation}\label{eq: cs2 in terms of A and B}
    c_s^2 = \frac{1}{1 - \frac{B(Y,\psi)}{A(Y,\psi)}} \,.
\end{equation}
Now, a necessary condition for stability is that $c_s^2\geq 0$, which in turn implies\footnote{The case $A(Y,\psi) = 0$ implies $c_s^2 = 0$, which trivially satisfies both stability and subluminality for all $B(Y,\psi)$. For this reason, the requirements we shall present here implicitly assume that we are in the case $A(Y,\psi)\neq 0$.} that
\begin{equation}
    \frac{B(Y,\psi)}{A(Y,\psi)} < 1 \,.
\end{equation}
Requiring also the avoidance of superluminal perturbations ($c_s^2 \leq 1$), one obtains
\begin{equation}
        \frac{B(Y,\psi)}{A(Y,\psi)} \leq 0 \,.
\end{equation}
These are the general expressions, with which one can verify that in the kinetic regime we recover the previous expressions for the stability \eqref{eq: stability in covar kinetic} and subluminal propagation \eqref{eq: subluminal in covar kinetic}. It is worth noting, however, that in the kinetic regime the constant $c_\rho$ cancels out rather early in the process (as early as in the EoS parameter, c.f. \eqref{eq: wcov}) and completely disappears from the subsequent treatment. However, when the potential is non-vanishing, it explicitly enters all of the analysis.

We shall now discuss the simple case of equal coupling functions, the GR limit, and finally translate our results into the TDiff frame.

\subsubsection{Equal coupling functions, $H_k = H_v \equiv H$} In the case where both coupling functions coincide, the mentioned expressions simplify significantly. In such a case, we obtain
\begin{equation}\label{eq: speed of sound constant potential equal coupling functions}
    c_s^2 = \frac{1}{1 + \frac{4 (H')^3}{H} \frac{X(Y,\psi)}{c_\rho H''}} \,.
\end{equation}
So, the stability of the perturbations requires
\begin{equation}
    \frac{(H')^3}{H} \frac{X(Y,\psi)}{c_\rho H''} > - \frac{1}{4} \,.
\end{equation}
Moreover, in order to avoid superluminalities, taking into account that $H>0$ and $X>0$, one also needs
\begin{equation}
    \frac{H'}{c_\rho H''} > 0 \,.
\end{equation}
Once again, these inequalities represent physically reasonable conditions which should help in selecting physically allowed coupling functions.

\subsubsection{GR limit} Let us now consider the GR limit for the quantities discussed in this section. Approaching this limit, $A(Y,\psi)$ and $B(Y,\psi)$ behave as
\begin{subequations}
    \begin{align}
        A(Y,\psi) &= (1+ \epsilon h) \left[ 0 - \frac{c_\rho}{2} \epsilon h'' \right] =\order{\epsilon^2} \,,\\[5pt]
        B(Y,\psi) &= 2 (\epsilon h')^2 \left( \epsilon h' V - \frac{c_\rho}{2} \right) = \order{\epsilon^3} \,,
    \end{align}
\end{subequations}
where we have recalled that $c_\rho = \order{\epsilon}$. The fraction then behaves as
\begin{equation}
    \frac{B(Y,\psi)}{A(Y,\psi)} = \order{\epsilon} \,,
\end{equation}
which vanishes in the limit $\epsilon \to 0$. As a result, in the GR limit we have that $c_s^2 \to 1$, as it should be for a Diff invariant scalar field with a canonical kinetic term (c.f. our results at the end of the $H_k = \text{const.}$ discussion, and see also reference \cite{Gordon:2004ez} for more details).

\subsubsection{Translation to the TDiff frame} We shall finally translate our general results to the TDiff frame. In the $\bar{\mu}=1$ frame, the energy density and pressure \eqref{eq: covar energy density and pressure} translate to
\begin{subequations}\label{eq: TDiff energy density and pressure}
    \begin{align}
        \left. \rho \right|_{\bar{\mu}=1} &= \frac{1}{\sqrt{g}} \left( f_k X + f_v V - \frac{c_\rho}{2} \right) \,,\\[5pt]
        \left. p \right|_{\bar{\mu}=1} &= \frac{1}{\sqrt{g}} \left( f_k X - f_v V + \frac{c_\rho}{2} \right) \,,
    \end{align}
\end{subequations}
which recalling the general constraint \eqref{eq: general metric constraint} can be seen to be equivalent to the previously presented energy density \eqref{eq: general energy density} and pressure \eqref{eq: general pressure}, respectively. The EoS parameter \eqref{eq: covar EoS general case} in the TDiff frame reads
\begin{equation}
    \left. w \right|_{\bar{\mu}=1} = \frac{f_k X - f_v V + c_\rho/2}{f_k X + f_v V - c_\rho/2} \,.
\end{equation}
Once again we may see that the fluid is non-adiabatic, as the general metric constraint \eqref{eq: general metric constraint} reveals that
\begin{equation}
    X = \frac{(f_v - 2g f_v') V(\psi) - c_\rho/2}{f_k - 2g f_k'} = X(g,\psi) \,,
\end{equation}
and so the energy density and pressure above are functions of both the metric determinant $g$ and the scalar field $\psi$.

On the other hand, expression \eqref{eq: cs2 in terms of A and B original} for $c_s^2$ evaluated in the TDiff frame may be written as
\begin{equation}
    \left. c_s^2 \right|_{\bar{\mu}=1} = \frac{a(g,\psi)}{a(g,\psi)- b(g,\psi)} \,,
\end{equation}
where we are denoting
\begin{subequations}
    \begin{align}
        \begin{split}
            a(g,\psi) &\equiv A(Y,\psi) \big|_{\bar{\mu}=1} = - c_\rho g f_k (f_k' + 2gf_k'') \\
            & \quad + 2g f_k V \bigg[ (f_k' + 2gf_k'')(f_v - 2g f_v') \\
            &\qquad \qquad \quad- (f_k - 2g f_k')(f_v' + 2gf_v'') \bigg] \,,
        \end{split}\\[10pt]
        b(g,\psi) &\equiv B(Y,\psi) \big|_{\bar{\mu}=1} = 2(f_k - 2g f_k')^3 X(g,\psi) \,.
    \end{align}
\end{subequations}
Once again, whenever $a(g,\psi)\neq 0$ we may find a common factor and write
\begin{equation}
    \left. c_s^2 \right|_{\bar{\mu}=1} = \frac{1}{1 - \frac{b(g,\psi)}{a(g,\psi)}} \,.
\end{equation}
The requirement of stable perturbations translates to
\begin{equation}
    \frac{b(g,\psi)}{a(g,\psi)} < 1 \,,
\end{equation}
while in order to also avoid superluminal propagations we must have
\begin{equation}
    \frac{b(g,\psi)}{a(g,\psi)} \leq 0 \,.
\end{equation}
Although these are the general conditions, we can once more find some simplifications when the coupling functions coincide ($f_k = f_v \equiv f$), which allow us to obtain the following requirements for stability
\begin{equation}
    \frac{(f-2gf')^3}{gf} \frac{X(g,\psi)}{c_\rho (f' + 2gf'')} > - \frac{1}{2}
\end{equation}
and subluminality
\begin{equation}
    \frac{f-2gf'}{c_\rho (f' + 2gf'')} > 0 \,,
\end{equation}
respectively. In any case, all of the conditions above should be helpful in deciding whether a particular TDiff model is physically reasonable. All in all, we have seen that covariantized treatment is quite direct, and the subsequent translation of the results to the TDiff framework fairly straightforward.

Thus concludes the most general situation of our non-adiabatic fluid. In the following section, we will discuss some particular cases of interest in which the fluid is adiabatic.

\subsection{Adiabatic models}\label{subsecVB}
We described in the previous section how the most general situation was a non-adiabatic fluid. In this section we will perform a more detailed discussion of adiabatic models. We begin by recalling equation \eqref{eq: general delta p} for $\delta p$, where we note that in order to have an adiabatic fluid the second term should vanish, meaning that the general condition for an adiabatic fluid is simply $\alpha = 0$. Now, $\alpha$ is itself a sum of two terms,
\begin{equation}\label{eq: alpha in two terms}
    \alpha = \underbrace{(-c_s^2) \left. \frac{\partial \rho}{\partial \psi} \right|_Y}_{\text{(i)}} + \underbrace{\left. \frac{\partial p}{\partial \psi} \right|_Y}_{\text{(ii)}} = 0 \,,
\end{equation}
so there are two possibilities: it could happen that the terms (i) and (ii) vanish separately, or it could happen that they do not vanish separately but their combination does. In the following we shall consider these possibilities in detail, and also study their implications.

\subsubsection{Case I}
We begin with the case in which the two terms in $\alpha$ vanish separately. More in particular, let us begin by studying the conditions under which the term $\text{(ii)} = 0$. Differentiating the pressure as it stands in equation \eqref{eq: covar energy density and pressure after substituting X(Y,psi)} we obtain
\begin{equation}
    \text{(ii)} = V' \left( \frac{H_k}{H_k'} H_v' - H_v \right) = 0 \,.
\end{equation}
Thus, term (ii) will vanish in situations in which $V'=0$ (i.e. $V=V_0$ a constant potential) and in situations in which
\begin{equation}
    \frac{H_k}{H_k'} H_v' - H_v = 0 \,.
\end{equation}
There are a couple of options at this point. If $H_v' = 0$ then the equation tells us that $H_v = 0$. If $H_v' \neq 0$ then we may rearrange the above expression as
\begin{equation}
    \frac{H_k'}{H_k} = \frac{H_v'}{H_v} \,,
\end{equation}
which may be straightforwardly integrated to yield
\begin{equation}
    H_v = C H_k \,,
\end{equation}
with $C$ a constant of integration. Thus, we conclude that there are three situations in which term (ii) vanishes:
\begin{itemize}
    \item A constant potential $V=V_0$.
    \item A vanishing potential coupling $H_v = 0$.
    \item A potential coupling of the form $H_v = C H_k$. 
\end{itemize}
Next we move on to the first term, and study the the conditions under which $\text{(i)} = 0$. Differentiating the energy density as it stands in equation \eqref{eq: covar energy density and pressure after substituting X(Y,psi)} we obtain
\begin{equation}
    \text{(i)} = (-c_s^2) \, V' \left( \frac{H_k}{H_k'} H_v' + H_v \right) = 0 \,.
\end{equation}
The simplest option is common with the previous case: term (i) will vanish when the potential $V=V_0$ is constant. Another possibility is that
\begin{equation}
    \frac{H_k}{H_k'} H_v' + H_v = 0 \,,
\end{equation}
which again admits the solution $H_v = 0$ and also
\begin{equation}
    \frac{H_k'}{H_k} = -\frac{H_v'}{H_v} \,.
\end{equation}
Integrating, we would find
\begin{equation}
    H_v = \frac{C}{H_k} \,,
\end{equation}
with $C$ a constant of integration. The final possibility is the vanishing of $c_s^2$ or, equivalently, the vanishing of $A(Y,\psi)$ as equation \eqref{eq: cs2 in terms of A and B original} reveals. This amounts to
\begin{equation}
    V \left( H_k'' H_v' - H_k' H_v'' \right) - \frac{c_\rho}{2} H_k'' = 0 \,.
\end{equation}
If the potential was identically zero, it would be nothing but a particular case of constant potential, and we already know that both terms (i) and (ii) would vanish. Thus, it would not be necessary to keep probing for information regarding the vanishing of $c_s^2$. In the following calculations, we shall assume that $V\neq 0$, so that we may rearrange to get
\begin{equation}
    H_k'' H_v' - H_k' H_v'' = \frac{c_\rho}{2V} H_k'' \,.
\end{equation}
At this point, it could happen that $H_k'' = 0$, which on the one hand implies $H_k = a Y + b$ and on the other
\begin{equation}
    H_v'' = 0
\end{equation}
(where we have recalled $H_k'\neq 0$), so that $H_v = c Y + d$. On the other hand, in cases where $H_k''\neq 0$ we could divide through and obtain the following:
\begin{equation}
    H_v' - \frac{H_k'}{H_k''} H_v'' = \frac{c_\rho}{2V} \,.
\end{equation}
We remark that the left-hand side of this equation is a function of only $Y$, while the right-hand side is a function of only $\psi$. The only way in which this is consistent is if both sides equal the same constant $K$. In particular, this means that we get a constant potential
\begin{equation}
    V = \frac{c_\rho}{2K} = V_0 \,,
\end{equation}
which takes us to the first case and so we do not need to look for further information. We thus conclude that, in practice, there are four situations in which term (i) vanishes:
\begin{itemize}
    \item A constant potential $V=V_0$.
    \item A vanishing potential coupling $H_v = 0$.
    \item A potential coupling of the form $H_v = \frac{C}{H_k}$.
    \item Both coupling functions are linear: $H_k = a Y + b$ and $H_v = c Y + d$.
\end{itemize}
Contrasting these four situations for which $\text{(i)}=0$ with the three for which $\text{(ii)}=0$, bearing in mind that both terms must vanish separately and simultaneously, reveals that in practice we can only have three cases in which both terms (i) and (ii) vanish independently:
\begin{enumerate}
    \item A constant potential $V=V_0$.
    \item A vanishing potential coupling $H_v = 0$.
    \item A linear kinetic coupling function $H_k = a Y + b$ and at the same time a potential coupling function of the form $H_v = C H_k$ (which will also be linear).
\end{enumerate}
The physical implications of each of these subcases shall be studied with greater detail later on. For now, let us remark that the reader may find in Table \ref{tab: adiabatic models} a summary of the results obtained.

\begin{table*}[t]
    \centering
    \def\arraystretch{2.5}
    \setlength\tabcolsep{2mm}
    \begin{tabular}{l|c|c|c|c|c}
        \multicolumn{1}{c|}{Name} & $V(\psi)$ & $H_k(Y)$ & $H_v(Y)$ & $c_\rho$ & restrictions \\ \hline\hline
        I.1. Shift symmetric model & --- & --- & 0 & --- & $H_k'\neq 0$ \\
        I.2. Constant potential model & $V_0$ & --- & --- & --- & $H_k'\neq 0$ \\
        II.1. Unstable model & --- & $e^{aY + b}$ & $d$ & --- & $a,d \neq 0$ \\
        II.2. Constant EoS model & --- & $a \,(H_v)^b$ & $H_v'\neq 0$ & 0 & $a,b\neq 0$ \\
        \makecell{II.3. Constant speed of sound model \\ ($b=1$ corresponds to subcase I.3.)} & --- & $a\left( Y + \frac{d}{c} \right)^b$ & $cY + d$ & --- &  $a,b,c\neq 0$ \\ \hline
        III. GR fluid model & $\frac{c_\rho}{2c}$ & $k$ & $cY + d$ & $\neq 0$ & $c\neq 0$
    \end{tabular}
    \caption{Summary of the six (independent) adiabatic TDiff models, where the first five correspond to $H_k'\neq 0$ and the last one to $H_k' = 0$. In the table, a long hyphen ``---'' means that the quantity in question is arbitrary, and the integration constants $\{ V_0,\, C,\, a,\, b,\, c,\, d, \,k\}$ are unrestricted unless otherwise specified.}
    \label{tab: adiabatic models}
\end{table*}

\subsubsection{Case II}
We now consider the case in which the two terms in $\alpha$ do not vanish separately but their combination does. Now, since none of the terms in equation \eqref{eq: alpha in two terms} for $\alpha$ vanishes individually, we can rearrange said expression to obtain:
\begin{equation}
    c_s^2 = \left. \frac{\partial p / \partial \psi}{\partial \rho / \partial \psi}\right|_Y \,.
\end{equation}
Differentiating the energy density and pressure in equations \eqref{eq: covar energy density and pressure after substituting X(Y,psi)} and recalling expression \eqref{eq: cs2 in terms of A and B original} for $c_s^2$ we find that
\begin{equation}\label{eq: adibaticity when combination vanishes}
    \frac{A(Y,\psi)}{A(Y,\psi) - B(Y,\psi)} = \frac{\frac{H_k}{H_k'} H_v' - H_v}{\frac{H_k}{H_k'} H_v' + H_v} \,.
\end{equation}
This would be the general relation which establishes the adiabaticity of the fluid in the case where both terms in $\alpha$ do not vanish separately but only through their combination.

In cases where $H_v'=0$ we would on the one hand have that $H_v = \text{const.}$ (different from zero, since $H_v = 0$ belongs to the two terms in $\alpha$ vanishing separately) and on the other it would follow from the above equation that
\begin{equation}
    c_s^2 = -1 \,.
\end{equation}
This would thus be an unstable adiabatic fluid. Despite this fact, we could carry further the analysis and find
\begin{equation}
    B(Y,\psi) = 2 A(Y,\psi) \,.
\end{equation}
Substituting expressions \eqref{eq: A(Y,psi) and B(Y,psi)}, using that $H_v'=0$, one obtains
\begin{equation}
    c_\rho (H_k')^2 = c_\rho H_k H_k'' \,,
\end{equation}
which, after cancelling a common $c_\rho$ (nonzero, since in our $H_v'=0$ study it would imply that $A(Y,\psi)=0$ and that is not the case), may be rearranged to find
\begin{equation}
    \frac{H_k'}{H_k} = \frac{H_k''}{H_k'} \,.
\end{equation}
Integrating twice, this would yield a kinetic coupling function of the form
\begin{equation}
    H_k = e^{aY + b} \,,
\end{equation}
with $a$ and $b$ constants of integration.

Consider now the cases where $H_v'\neq 0$. This means that we could find a common factor in the right-hand side of equation \eqref{eq: adibaticity when combination vanishes} and write it as
\begin{equation}
    \frac{1}{1-\frac{B(Y,\psi)}{A(Y,\psi)}} = \frac{\frac{H_k}{H_k'} - \frac{H_v}{H_v'}}{\frac{H_k}{H_k'} + \frac{H_v}{H_v'}} \,,
\end{equation}
where we have also simplified the left-hand side recalling that $A(Y,\psi) \neq 0$ (if it vanished we would be in a particular case of the previous study). We can rearrange this equation to find
\begin{equation}\label{eq: varphi appears}
    \frac{B(Y,\psi)}{A(Y,\psi)} = 1 - \frac{\frac{H_k}{H_k'} + \frac{H_v}{H_v'}}{\frac{H_k}{H_k'} - \frac{H_v}{H_v'}} \equiv \varphi(Y) \,.
\end{equation}
(Note that in the situation under study $\varphi(Y)\neq 0$, because its vanishing would imply $H_v = 0$ as follows from the above definition of $\varphi(Y)$, and this is incompatible with $H_v'\neq 0$). Now, the right-hand side is a function only of $Y$, while the left-hand side depends on both $Y$ and $\psi$. Let us try to localize better these dependencies and see what we obtain. We first of all write the equation as
\begin{equation}
    B(Y,\psi) = \varphi(Y) A(Y,\psi) \,.
\end{equation}
Substituting now the expressions for $A(Y,\psi)$ and $B(Y,\psi)$ from equation \eqref{eq: A(Y,psi) and B(Y,psi)} and rearranging the result to group terms with $V$ on one side, we obtain
\begin{equation}
    \begin{split}
        V&\left\{ 2(H_k')^2 H_v' - \varphi(Y) H_k \left( H_k'' H_v' - H_k' H_v'' \right) \right\} = \\[5pt]
        &\qquad\qquad= \frac{c_\rho}{2} \left[ 2(H_k')^2 - \varphi(Y) H_k H_k'' \right] \,.
    \end{split}
\end{equation}
Now, if the term inside curly brackets was non-zero, it would mean that we could solve for the potential $V(\psi)$ and it would be a function of only $Y$. The only way this is compatible is if both sides were constant, but a constant potential is part of the previous situation and does not belong to the present study. Hence, it must be the case that the term in between curly brackets in the left-hand side of the above equation vanishes, implying in turn that the full right-hand side must vanish as well. We thus conclude:
\begin{subequations}
    \begin{align}
        2(H_k')^2 H_v' - \varphi(Y) H_k \left( H_k'' H_v' - H_k' H_v'' \right) = 0 \,,\\[5pt]
        \frac{c_\rho}{2} \left[ 2(H_k')^2 - \varphi(Y) H_k H_k'' \right] = 0 \,.
    \end{align}
\end{subequations}
Let us focus on the first expression. Dividing by the nonzero $H_v'$ and rearranging, we find that
\begin{equation}\label{eq: simplify 1st expression}
    2(H_k')^2 - \varphi(Y) H_k H_k'' = - \varphi(Y) H_k H_k' \frac{H_v''}{H_v'} \,.
\end{equation}
Substituting this result into the second expression, we obtain
\begin{equation}
    -\frac{c_\rho}{2} \varphi(Y) H_k H_k' \frac{H_v''}{H_v'} = 0 \,,
\end{equation}
which implies
\begin{equation}
    c_\rho \, H_v'' = 0 \,.
\end{equation}
Thus, the two possibilities for the cases where $H_v'\neq 0$ are $c_\rho = 0$ and $H_v'' = 0$. The analysis for $c_\rho=0$ deserves greater care, but regarding $H_v''=0$ we can already say that (since $H_v'\neq 0$) we must have $H_v = cY + d$ with $c\neq 0$ and $d$ constants of integration.

Let us now pause for a moment and make a list of the three distinct situations in which the two terms in $\alpha$ do not vanish identically but their combination does:
\begin{itemize}
    \item $H_v = v \neq 0$ and $H_k = e^{aY + b}$ (unstable model).
    \item $H_v'\neq0$ and $c_\rho = 0$.
    \item $H_v = cY + d$, with $c\neq 0$.
\end{itemize}
The first case yields an unstable model, as we have discussed. Regarding the other two situations, we still do not know what the form of the kinetic coupling function $H_k$ is, so let us focus on that now.

The case in which the constant $c_\rho$ vanishes (while $H_v'\neq0$) turns out to be quite simple, as many convenient cancellations take place. Indeed, evaluating $A(Y,\psi)$ and $B(Y,\psi)$ from equation \eqref{eq: A(Y,psi) and B(Y,psi)} whenever $c_\rho=0$ results in
\begin{equation}
    \frac{B(Y,\psi)}{A(Y,\psi)} = 2 \frac{\frac{H_k'}{H_k}}{\frac{H_k''}{H_k'}- \frac{H_v''}{H_v'}} \,,
\end{equation}
and we see that the potential $V$ completely disappears from the ratio. Referring back to equation \eqref{eq: varphi appears}, it follows that
\begin{equation}
    2 \frac{\frac{H_k'}{H_k}}{\frac{H_k''}{H_k'}- \frac{H_v''}{H_v'}} = (-2) \frac{\frac{H_v}{H_v'}}{\frac{H_k}{H_k'} - \frac{H_v}{H_v'}} \,.
\end{equation}
This equation may be simplified and rearranged carefully to find
\begin{equation}
    \frac{H_k''}{H_k'} - \frac{H_k'}{H_k} = \frac{H_v''}{H_v'} - \frac{H_v'}{H_v} \,.
\end{equation}
Integrating once, it follows that
\begin{equation}
    \frac{H_k'}{H_k} = b \frac{H_v'}{H_v} \,,
\end{equation}
with $b$ an integration constant (nonzero, recall that all throughout $H_k'\neq 0$). Integrating a second time, we finally conclude that
\begin{equation}
    H_k = a (H_v)^b \,,
\end{equation}
with $a$ another (nonzero) integration constant. In this way, given any potential coupling function $H_v$ which satisfies $H_v'\neq 0$ and a kinetic coupling function of the above form, the fluid will be adiabatic.

Now we move on to the other option, i.e. a linear potential coupling function $H_v = cY + d$, with $c\neq 0$. Substituting this form into expression \eqref{eq: simplify 1st expression} we find the following equation for the kinetic coupling function:
\begin{equation}
    2 (H_k')^2 = \varphi(Y) H_k H_k'' =  (-2)\frac{Y+d/c}{\frac{H_k}{H_k'} - (Y + d/c)} \, H_k H_k'' \,,
\end{equation}
where in the second equality we have used the definition \eqref{eq: varphi appears} for $\varphi(Y)$ particularized to our case. The above expression may be simplified and rearranged to find
\begin{equation}
    \frac{H_k'}{H_k} - \frac{H_k''}{H_k'} = \frac{1}{Y + d/c} \,.
\end{equation}
Integrating once, it follows that
\begin{equation}
    \frac{H_k'}{H_k} = \frac{b}{Y + d/c} \,,
\end{equation}
with $b$ a (nonzero) integration constant, and integrating a second time one may finally obtain
\begin{equation}
    H_k = a (Y+d/c)^b \,,
\end{equation}
with $a$ another (nonzero) integration constant.

Thus, the three possibilities for adiabatic models that we find in Case II would be:
\begin{enumerate}
    \item $H_v = d \neq 0$ and $H_k = e^{aY + b}$ (unstable model).
    \item $c_\rho = 0$, with $H_v'\neq0$ and $H_k = a (H_v)^b$.
    \item $H_v = cY + d$ and $H_k = a (Y+d/c)^b$.
\end{enumerate}
Thus concludes the possible ways in which the models may be adiabatic. For simplicity and an easy reference, we have also included these results in Table \ref{tab: adiabatic models}.

We now move on to the analysis of the physical implications of each of the six adiabatic subcases we have found.

\subsubsection*{Subcase I.1. -- Shift symmetric model}
We begin by noting that the particular model with $H_v = 0$ is shift symmetric, and its results are fundamentally those found in the kinetic domination regime (which we already knew to be adiabatic). If we compute the energy density, pressure, EoS parameter, and adiabatic speed of sound, we straightforwardly reobtain the same results as those presented in section \ref{subsecIVB}. Since this case has been previously analyzed in detail, we shall not reproduce it again here.

\subsubsection*{Subcase I.2. -- Constant potential model}
Let us continue by considering the case in which we have both kinetic and potential contributions but the potential is $V = V_0 = \text{const.}$ The general EoM \eqref{eq: covar EoM psi} for $\psi$ actually reduces to the kinetic EoM \eqref{eq: covar EoM psi kinetic}, whose solution is known and is given by \eqref{eq: covar EoM psi kinetic solutions}. Substituting into the solutions \eqref{eq: EoM Tmu solutions} for the EoM of $T^\mu$, we obtain
\begin{equation}
    \frac{H_k'}{2} \, \frac{C_\psi(x)}{(H_k \delta V)^2} - H_v' V_0 = -\frac{c_\rho}{2} \,,
\end{equation}
and going to the $\bar{\mu}=1$ frame it follows that
\begin{equation}
    \frac{(f_k - 2gf_k')}{2} \, \frac{g \, C_\psi(x)}{(f_k \delta V)^2} - (f_v - 2gf_v') \, V_0 = -\frac{c_\rho}{2} \,.
\end{equation}
This expression is the general constraint on the metric whenever we have a kinetic term and a constant potential.

Now, the energy density and pressure are found from \eqref{eq: covar energy density and pressure after substituting X(Y,psi)} to be
\begin{subequations}
    \begin{align}
        \rho &= \frac{H_k}{H_k'} \left( H_v' V_0 - \frac{c_\rho}{2} \right) + H_v V_0 - \frac{c_\rho}{2} Y = \rho(Y) \,,\\[5pt]
        p &= \frac{H_k}{H_k'} \left( H_v' V_0 - \frac{c_\rho}{2} \right) - H_v V_0 + \frac{c_\rho}{2} Y = p(Y)\,.
    \end{align}
\end{subequations}
Their dependence on a single variable reveals that we are dealing with an adiabatic fluid, with EoS parameter
\begin{equation}\label{eq: EoS constant potential}
    w = \frac{c_\rho \left( H_k - Y H_k' \right) + 2 V_0 \left( H_v H_k' - H_k H_v' \right)}{c_\rho \left( H_k + Y H_k' \right) - 2 V_0 \left( H_v H_k' + H_k H_v' \right)} \,.
\end{equation}
We may also calculate the speed of sound of the adiabatic perturbations using relation \eqref{eq: speed of sound of adiabatic perturbations}, which yields
\begin{equation}\label{eq: covar speed of sound with V0}
    c_a^2 = \frac{A(Y)}{A(Y) - B(Y)} \,,
\end{equation}
where we are denoting
\begin{subequations}\label{eq: A(Y) and B(Y)}
    \begin{align}
        A(Y) &\equiv V_0 H_k \left( H_k'' H_v' - H_k' H_v'' \right) - \frac{c_\rho}{2} H_k H_k'' \,,\\[5pt]
        B(Y) &\equiv 2(H_k')^2 \left( H_v' V_0 - \frac{c_\rho}{2} \right) \,.
    \end{align}
\end{subequations}
Before proceeding any further, let us stress a couple of points. Firstly, the above expression \eqref{eq: covar speed of sound with V0} reduces to the adiabatic speed of sound \eqref{eq: cacov v1} in the kinetic regime, as one would expect. Secondly, and perhaps more interestingly, let us remark that we have obtained equation \eqref{eq: covar speed of sound with V0} for the adiabatic speed of sound via the relation \eqref{eq: speed of sound of adiabatic perturbations} (valid precisely for adiabatic perturbations). Nevertheless, in the case of a constant potential it is immediate to see that the $\psi$ dependence in the quantities of the previous section completely disappears: from equations \eqref{eq: A(Y,psi) and B(Y,psi)} and \eqref{eq: A(Y) and B(Y)} we see that $\left. A(Y,\psi) \right|_{V_0} = A(Y)$ and $\left. B(Y,\psi) \right|_{V_0} = B(Y)$, while from \eqref{eq: cs2 in terms of A and B} and \eqref{eq: covar speed of sound with V0} we verify that $\left. c_s^2 \right|_{V_0} = c_a^2$. So, consistently, this result could have been obtained noting that for a constant potential we would have $\alpha=0$ in equation \eqref{eq: general delta p}, as we previously mentioned.

Finally, requiring the stability of the perturbations translates to $B(Y)/A(Y) < 1$, and if we also wish to have subluminality then it must be the case that $B(Y)/A(Y) \leq 0$.

All in all, the case of a constant potential is a simple but non-trivial example which the TDiff approach was unable to reach due to the difficulty related with the study of EMT conservation, but which the covariantized approach allows to study in a straightforward manner.

\subsubsection*{Subcase I.3. -- Constant pressure model}
Another case in which the fluid is adiabatic is the one in which the two couplings are linear and related by
\begin{subequations}
    \begin{align}
        H_k &= a Y + b \,,\\[5pt]
        H_v &= C H_k \,.
    \end{align}
\end{subequations}
The energy density and pressure are found from \eqref{eq: covar energy density and pressure after substituting X(Y,psi)} to be
\begin{subequations}
    \begin{align}
        \rho &= 2C(a Y + b) V - \frac{c_\rho}{2}\left(2Y+\frac{b}{a}\right) \,,\\[5pt]
        p &= - \frac{c_\rho b}{2 a} = \text{constant} \,.
    \end{align}
\end{subequations}
Looking at the above quantities, a couple of features immediately jump out. The first one is that the pressure is constant, and as a result we find that the fluid will have no pressure perturbations, i.e. $\delta p = 0$. If we compute the speed of sound \eqref{eq: cs2 in terms of A and B original} for this model, it consistently reveals that $c_s^2 = 0$ (satisfying both stability and subluminality).

The second feature is that the energy density depends in general on both $Y$ and $\psi$, and so one could wonder how the adiabatic speed of sound could be calculated given that we have a dependence on two variables here. This is solved by again remarking that the pressure is constant, so that from the definition of the adiabatic speed of sound as
\begin{equation}
    c_a^2 = \frac{\dot{p}}{\dot{\rho}} = \frac{u^\alpha \partial_\alpha p}{u^\beta \partial_\beta \rho}
\end{equation}
we consistently find $c_a^2 = 0 = c_s^2$, and the two coincide (as they should for our adiabatic fluid).

\subsubsection*{Subcase II.1. -- Unstable model}
Although the model defined by the coupling functions
\begin{subequations}
    \begin{align}
        H_k &= e^{aY + b} \,, \\[5pt]
        H_v &= d \,,
    \end{align}
\end{subequations}
yields unstable perturbations, we shall nevertheless discuss it for completeness. The energy density and pressure follow from \eqref{eq: covar energy density and pressure after substituting X(Y,psi)} and read
\begin{subequations}
    \begin{align}
        \rho &= -\frac{c_\rho}{2}\left(\frac{1}{a} + Y\right) + Vd \,,\\[5pt]
        p &= -\frac{c_\rho}{2}\left(\frac{1}{a} - Y\right) - Vd  \,.
    \end{align}
\end{subequations}
Once again they depend on a couple of variables, and so the correct way of finding the adiabatic speed of sound is from its definition as
\begin{equation}
    c_a^2 = \frac{\dot{p}}{\dot{\rho}} = -1 \,,
\end{equation}
an unstable result which we expected, coinciding with the speed of sound $c_s^2 = -1$ as obtained from \eqref{eq: cs2 in terms of A and B original}.

\subsubsection*{Subcase II.2. -- Constant EoS model}
The main feature of this subcase is the vanishing of the constant $c_\rho$ (which greatly simplifies the treatment) and we recall that the coupling functions are related by
\begin{equation}
    H_k = a (H_v)^b \,,
\end{equation}
such that $a,b,H_v'\neq 0$. The energy density and pressure are found substituting in \eqref{eq: covar energy density and pressure after substituting X(Y,psi)}, and take the form
\begin{subequations}
    \begin{align}
        \rho &= \frac{1+b}{b} \, H_v V \,,\\[5pt]
        p &= \frac{1-b}{b} \, H_v V  \,,
    \end{align}
\end{subequations}
and the EoS parameter turns out to be constant:
\begin{equation}
    w = \frac{1-b}{1+b} \,.
\end{equation}
Finally, the speed of sound \eqref{eq: cs2 in terms of A and B original} reads
\begin{equation}\label{eq: ca2 in II2}
    c_s^2 = \frac{1-b}{1+b} \,,
\end{equation}
and it indeed coincides with the adiabatic speed of sound since, for the case of a constant EoS parameter, we have $c_a^2 = w = c_s^2$. The requirement of stability translates to the possibilities
\begin{equation}\label{eq: stability II2}
    -1 < b < 0 \quad \text{\&} \quad 0 < b \leq 1 \,,
\end{equation}
and also requiring subluminality gives
\begin{equation}\label{eq: subluminality II2}
    b > 0 \,.
\end{equation}
In obtaining the above results we have recalled that we are in a model in which $b\neq 0$ from the start, so we must remove the possibility $b=0$ from them.

\subsubsection*{Subcase II.3. -- Constant speed of sound model}
Another particular model yielding an adiabatic fluid is that in which the coupling functions take the form:
\begin{subequations}
    \begin{align}
        H_k &= a \left(Y+\frac{d}{c}\right)^b \,, \\[5pt]
        H_v &= cY + d \,,
    \end{align}
\end{subequations}
with all constants different from zero except perhaps $d$. A trivial rewriting of the kinetic coupling function as
\begin{equation}
    H_k = a c^b (cY + d)^b= a c^b (H_v)^b
\end{equation}
reveals that this case is closely related to the previous one, where now $H_v$ is linear, but with the drawback that now the constant $c_\rho$ does not necessarily vanish. If the initial conditions where chosen such that $c_\rho = 0$, however, we would be in a particular realization of the previous case. Now, our choice of initial conditions affects the particular value of our fluid quantities (energy density, pressure, EoS) at any given time, but it should not affect the perturbations, and this is what we shall see.

The energy density and pressure for this model are found from \eqref{eq: covar energy density and pressure after substituting X(Y,psi)} to be
\begin{subequations}
    \begin{align}
        \rho &= \frac{1+b}{b} \left( -\frac{c_\rho}{2} Y + Vd + c Y V \right) - \frac{c_\rho d}{2bc} \,,\\[5pt]
        p &= \frac{1-b}{b} \left( -\frac{c_\rho}{2} Y + Vd + c Y V \right) - \frac{c_\rho d}{2bc} \,,
    \end{align}
\end{subequations}
and the EoS parameter takes the form
\begin{equation}
    w = \frac{(1-b) \left( -\frac{c_\rho}{2} Y + Vd + c Y V \right) - \frac{c_\rho d}{2c}}{(1+b) \left( -\frac{c_\rho}{2} Y + Vd + c Y V \right) - \frac{c_\rho d}{2c}} \,.
\end{equation}
All of these fluid quantities would reduce to the ones in the previous case when the initial conditions were chosen such that $c_\rho = 0$. Now, both the energy density and pressure depend on two variables, and so in order to find the adiabatic speed of sound we must resort to the following definition:
\begin{equation}\label{eq: ca2 in II3}
    c_a^2 = \frac{\dot{p}}{\dot{\rho}} = \frac{1-b}{1+b} \,.
\end{equation}
Calculating the speed of sound from equation \eqref{eq: cs2 in terms of A and B original} yields the same result, so that indeed $c_s^2 = c_a^2$. Moreover, we may immediately see that the the adiabatic speed of sound \eqref{eq: ca2 in II3} is identical to that from the previous section, given in \eqref{eq: ca2 in II2}. So, indeed, even though our choice of initial conditions affects the ``background physics'' of our fluid, it does not affect its perturbations. Also, since the form of the adiabatic speed of sound is the same, we already know the requirements for stability and subluminality: they are \eqref{eq: stability II2} and \eqref{eq: subluminality II2}, respectively.

As a final note, the reader may recognize that the particular model given by $b=1$ (which means that the two coupling functions are linear and proportional to each other) reduces to the constant pressure situation I.3. we previously studied. Therefore, in a way, the present model is more general as it encompasses the aforementioned one.

With this last case, we have finally finished the discussion of the six situations in which the perturbations of our (generally non-adiabatic) fluid became adiabatic. Nevertheless, as we anticipated, we are still missing the analysis of the situations in which the kinetic coupling function $H_k$ is constant, so that $H_k' = 0$. For completeness, this detail shall be studied in the following section.

\subsection{Constant kinetic coupling function}\label{subsecVC}
After all of our general discussion, we come back to a question which remains open. Let us consider the case in which the potential coupling function is a constant, i.e.
\begin{equation}
    H_k = k = \text{const.}
\end{equation}
What happens in this situation? To begin with, since $H_k'=0$, the solutions \eqref{eq: EoM Tmu solutions} for the EoM of $T^\mu$ tell us that
\begin{equation}
    H_v' V = \frac{c_\rho}{2} \,.
\end{equation}
Possible solutions to this equation are $V=0$ and $H_v' = 0$, which both imply that $c_\rho = 0$. Beyond these trivial cases, we may divide through by the potential and find
\begin{equation}
    H_v' = \frac{c_\rho}{2V} \,.
\end{equation}
Now, the left-hand side depends only on $Y$ and the right-hand side only on $\psi$, and this is only possible if both sides equal the same constant $c$. On the one hand, this tells us that the potential takes on the constant value
\begin{equation}
    V = \frac{c_\rho}{2c} \,,
\end{equation}
and on the other that the potential coupling is linear,
\begin{equation}
    H_v = cY + d \,,
\end{equation}
with $d$ an integration constant. In summary, then, we in principle have three situations to study:
\begin{enumerate}
    \item $V = \frac{c_\rho}{2c} \neq 0$ and $H_v = cY + d$, with $c \neq 0$.
    \item $V=0$, with $H_v'$ arbitrary.
    \item $H_v' = 0$ ($\Rightarrow H_v = v =$ const.), with $V$ arbitrary.
\end{enumerate}
A couple of comments are due at this point. On the one hand, the second model gives simply
\begin{equation}
    S_\text{Diff} = S_\text{EH} + \int d^4x\,\sqrt{g}\, k X \,,
\end{equation}
and this is nothing but the standard GR case of a shift-symmetric scalar field. On the other hand, the third model gives
\begin{equation}
    S_\text{Diff} = S_\text{EH} + \int d^4x\,\sqrt{g} \left( k X - v V \right) \,,
\end{equation}
and this is nothing but the standard GR case of a general scalar field. Thus, since the second and third models are nothing but GR, we shall not study them in detail here, but rather we refer the reader to references \cite{Hu:1998kj,Gordon:2004ez}. Instead, we shall focus on the first case, which is the one that yields ``true'' TDiff results. Its analysis is done in the following subsection, and the reason for the numbering we have chosen will be understood once it is finished.

\subsubsection*{Subcase III -- GR fluid model}
Substituting the particular expressions $V = \frac{c_\rho}{2c}$ and $H_v = cY + d$ into equations \eqref{eq: covar energy density and pressure}, one finds some convenient cancellations which yield
\begin{subequations}
    \begin{align}
        \rho &= k X + \frac{c_\rho d}{2c} = \rho(X) \,,\\[5pt]
        p &= k X - \frac{c_\rho d}{2c} = p(X) \,,
    \end{align}
\end{subequations}
and the fluid is thus adiabatic once again. It is interesting to remark that, recalling the form of the potential, the above expressions may be rewritten as
\begin{subequations}
    \begin{align}
        \rho &= k X + V d \,,\\[5pt]
        p &= k X - V d \,,
    \end{align}
\end{subequations}
and so the energy density and pressure would fundamentally be identical to those of a canonical scalar field with constant potential in GR (modulo superfluous renormalizations of $\psi$). In this manner, even though our theory is not GR, we would obtain the same phenomenology. In particular, the EoS parameter reads
\begin{equation}
    w = \frac{k X - Vd}{k X + Vd} \,,
\end{equation}
which is the typical form in GR, and interpolates between $w=-1$ (potential domination) and $w=1$ (kinetic domination) Moreover, if we compute the adiabatic speed of sound from \eqref{eq: speed of sound of adiabatic perturbations} we obtain $c_a^2 = 1$, as is the case in GR.

We have thus studied the situation whenever the kinetic coupling function $H_k$ is constant, and this completes any of the questions left open in the previous analysis (which was carried out assuming $H_k' \neq 0$). Since we have obtained an adiabatic fluid, we have numbered this case as III and we have nicely included it in Table \ref{tab: adiabatic models} for an easy reference.

\subsection{Particular Diff solution in TDiff theories}\label{subsecVD}
As a final comment, we shall discuss a family of general TDiff models which can yield Diff solutions to the theory. A word of caution: we will perform the study in the covariantized approach, where strictly speaking the theories are (by construction) Diff invariant already; hence, whenever we speak of a ``TDiff theory'' in this context, we fundamentally mean that the coupling functions $H(Y)$ are not constant (meaning that the theory when going to the TDiff frame was not already Diff invariant to start with, or in other words, that we actually needed to introduce a new field to restore the symmetry).

Having cleared that up, let us consider the family of TDiff theories in which the coupling functions have a common extremum at $Y=Y_0$, i.e.
\begin{equation}\label{eq: extremum condition}
    H_k'(Y_0) = 0 = H_v'(Y_0) \,.
\end{equation}
Now, the solution $Y = Y_0 = \text{const.}$ is a valid (and trivial) solution to the EoM \eqref{eq: EoM Tmu} of the vector field $T^\mu$. Substituting now this solution into the EoM \eqref{eq: covar EoM psi} for $\psi$ yields
\begin{equation}
    H_k(Y_0)  \nabla_\mu \nabla^\mu\psi + H_v(Y_0) V'(\psi) = 0 \,.
\end{equation}
In this way, up to superfluous constant factors, the scalar field $\psi$ follows the usual Diff equation. Moreover, the EMT \eqref{eq: covariantized EMT} on these solutions reads
\begin{equation}
    T_{\mu\nu} = H_k(Y_0) \partial_\mu \psi \partial_\nu \psi - \left[ H_k(Y_0) X - H_v(Y_0) V \right] g_{\mu\nu} \,,
\end{equation}
and so we also recover the standard expression for a Diff theory with a scalar field (up to constant factors).

We stress that we are not setting the coupling functions $H(Y)$ to be constant from the start, which indeed amounts to the assertion that the original theory was Diff invariant already. Instead, we are working with non-constant, arbitrary coupling functions $H(Y)$ such that they have an extremum at $Y=Y_0$. These TDiff theories are explicitly not GR, but what we are seeing is that they admit a particular solution which reproduces the same results as GR. Thus, in principle, we would not be able to distinguish GR from a particular solution of a TDiff theory where the coupling functions reach an extremum at the same point. A valid and useful question in order to discern this would be: are these solutions stable under perturbations? By this we mean the following: were we to slightly perturb the solutions ($Y = Y_0 + \delta Y$, etc.), would the perturbation $\delta Y$ decay so that we would reach again the $Y=Y_0$ behaviour? We shall in the following address this question.

\subsubsection{Perturbations}
Let us begin by considering the evolution of the perturbations in general, before particularizing to our family of theories. The EoM \eqref{eq: EoM Tmu} for $T^\mu$ may be written as
\begin{equation}
    \partial_\mu \mathcal{S}(X,Y,\psi) = 0 \,,
\end{equation}
where for simplicity we denote
\begin{equation}
    \mathcal{S}(X,Y,\psi) \equiv H_k'(Y) X - H_v'(Y) V(\psi) \,.
\end{equation}
The solutions \eqref{eq: EoM Tmu solutions} are the level surfaces
\begin{equation}
    \mathcal{S}(X,Y,\psi) = \text{const.} \equiv -\frac{c_\rho}{2} \,,
\end{equation}
where the particular constant (the particular ``level'') is fixed through the initial conditions on the fields and their derivatives. Now, suppose  we have found a solution $\bar{\Phi} \equiv \{\bar{X},\bar{Y},\bar{\psi}\}$ to the theory (which we shall refer to as our ``background''). Slightly perturbing it and demanding that the perturbed fields $\Phi = \bar{\Phi} + \delta \Phi$ are also a solution to the theory reveals that, in general, we will change what level surface our new (perturbed) solution lives on. More in particular, demanding
\begin{equation}
    \partial_\mu \mathcal{S}(\Phi) = \partial_\mu \mathcal{S}(\bar{\Phi} + \delta\Phi) = 0
\end{equation}
yields the following equation for the perturbations:
\begin{equation}\label{eq: general perturbations}
    \begin{split}
        H_k'(\bar{Y}) \delta X + \mathcal{B}(\bar{X},\bar{Y},\bar{\psi}) \delta Y - H_v'(\bar{Y}) V'(\bar{\psi}) \delta \psi =\\[5pt]
        = \text{const.} \equiv -\frac{\delta c_\rho}{2} \,,
    \end{split}
\end{equation}
where, to alleviate the view, we are defining the background quantity
\begin{equation}
    \mathcal{B}(\bar{X},\bar{Y},\bar{\psi}) \equiv H_k''(\bar{Y}) \bar{X} - H_v''(\bar{Y}) V(\bar{\psi}) \,.
\end{equation}
In the above equation, the constant $\delta c_\rho$ would now be fixed through the initial conditions on the perturbed fields $\Phi = \bar{\Phi} + \delta\Phi$ and their derivatives.

Note that this  would not be the only equation for the perturbations: we would also have the equations coming from perturbing the EoM of the scalar field $\psi$ and the Einstein equations. Nevertheless, in order to keep it brief, it will suffice to simply consider this one.

\subsubsection{Stability of the Diff solution in TDiff theories}
Having presented the general case, let us now consider the family of TDiff theories for which the coupling functions $H(Y)$ reach a common extremum at $H_k'(Y_0) = 0 = H_v'(Y_0)$. As we saw, the solution $\bar{Y} = Y_0 = \text{const.}$ is a valid solution to the EoM of $T^\mu$ (it is a trivial one in fact) which yields a behaviour identical to GR. Now, is this particular solution stable? Let us consider equation \eqref{eq: general perturbations} above, particularized to these theories. The extremum condition helps simplifying it down to
\begin{equation}
    \mathcal{B}(\bar{X},Y_0,\bar{\psi}) \delta Y = -\frac{\delta c_\rho}{2} \,.
\end{equation}
Assuming that the background quantity $\mathcal{B}(\bar{X},Y_0,\bar{\psi})$ is nonzero\footnote{If it vanished, it would on the one hand imply $\delta c_\rho = 0$ (meaning we do not change level surface), and on the other the equation would turn into a $0=0$ identity from which no further information could be extracted. As a result, we would have to consider the perturbed Einstein equations and the perturbed equations for the scalar field to extract some information.} and solving for the perturbation, we find that
\begin{equation}
    \delta Y = \frac{-\delta c_\rho/2}{\mathcal{B}(\bar{X},Y_0,\bar{\psi})} \,.
\end{equation}
In the particular situation in which we choose the initial conditions $\delta Y |_\text{initial} = 0$ (essentially meaning that the perturbed solution crosses $Y = Y_0$ at some point, which we take as the ``initial'' one), then $\delta c_\rho = 0$ and the perturbation $\delta Y$ would always be zero. Such a theory would thus be ``stable'' in the sense that the perturbation decays (it actually stays zero constantly) and we recover the initial situation of $Y=Y_0$. Nevertheless, in general, we will be in a situation in which $\delta c_\rho \neq 0$, and so if we wish to shed some light on the stability of our solution we must study the evolution of the background quantity $\mathcal{B}(\bar{X},Y_0,\bar{\psi})$. Since this is difficult to tackle in general, let us consider the two limiting regimes of potential and kinetic domination and see if we can gain some intuition about what might happen.

\paragraph*{Potential domination.}
In this case, the background quantity reads $\mathcal{B} = - H_v''(Y_0) V(\bar{\psi})$. Now, the EoM for $\psi$ in the potential regime tell us that the scalar field takes on the constant value $\bar{\psi} = \psi_0$ such that the potential reaches an extremum. As a result,
\begin{equation}
    \mathcal{B} = - H_v''(Y_0) V(\psi_0) = \text{const.}
\end{equation}
Therefore, the perturbation stays constant as well\footnote{This is actually a general result for the perturbation $\delta Y$ in the potential regime, not limited to the particular family of theories we are studying. This may be seen by evaluating equation \eqref{eq: general perturbations} in the potential regime.}:
\begin{equation}
    \delta Y = \text{constant} \,.
\end{equation}
This means that the solution $\bar{Y} = Y_0 = \text{const.}$ is \textit{not stable} in the potential regime, since the perturbation does not decay but stays frozen.

\paragraph*{Kinetic domination.}
In this case, the background quantity reads $\mathcal{B} = H_k''(Y_0) \bar{X}$. Now, from equation \eqref{eq: covar EoM psi kinetic solutions}, the EoM for $\psi$ in the kinetic regime tells us that
\begin{equation}
    2\bar{X} = \frac{C_\psi(x)}{(H_k(Y_0) \delta V)^2} \,.
\end{equation}
As a result, the background quantity behaves as
\begin{equation}
    \mathcal{B} = \frac{H_k''(Y_0) C_\psi(x)}{2 H_k^2(Y_0) (\delta V)^2} \,.
\end{equation}
Substituting this result back into the expression for the perturbation, we find that
\begin{equation}
    \delta Y = \frac{-\delta c_\rho H_k^2(Y_0)}{H_k''(Y_0) C_\psi(x)} \,(\delta V)^2 \equiv C_Y(x)\,(\delta V)^2 \,,
\end{equation}
where we have grouped the term multiplying $(\delta V)^2$ into $C_Y(x)$ such that $\dot{C}_Y(x) = u^\mu \partial_\mu C_Y(x) = 0$ (i.e. a constant in what respects evolution along the field's direction). In a cosmological scenario we have $\delta V = a^3$ and $C_Y = \text{constant}$, so the perturbation grows as $\delta Y \propto a^6$. This rapid growth with the scale factor reveals that the particular solution $Y=Y_0 = \text{const.}$ is \textit{not stable in a cosmological scenario} (at least in models without recollapse). Without directly going to the cosmological scenario, one can see from the above equation that the stability of the solution is related to the background cross-sectional volume diminishing along the scalar field's direction. We can make this statement more mathematically precise by taking the dot derivative:
\begin{equation}
    (\delta Y)^{\bm{\cdot}} = 2\,\delta Y \,(\ln \delta V)^{\bm{\cdot}} \,,
\end{equation}
which, recalling equation \eqref{eq: cross-sectional volume}, reveals that it is proportional to the expansion. In this way, we confirm that for a model with a positive expansion the perturbation will grow (making the solution unstable), whereas for a model with a negative expansion it will decay (making the solution stable).

As a final note, let us remark that even though in Subcase III from the previous section we also obtained a GR behaviour, in that case said behaviour is general for any solution to the equations of motion. Indeed, in Subcase III we are actually fixing the coupling functions and hence the theory, so all solutions of the theory will have that behaviour (we do not have any ``particular solution whose stability we should study''). In the present case, however, we are dealing with a particular solution of a family of theories, without actually fixing the coupling functions.

\section{Conclusions}\label{section: Conclusions}
In this work we have explored the idea of restoring the Diff invariance to a theory in which it was previously broken down to TDiff. Inspired by the work of Stueckelberg \cite{Stueckelberg} for gauge fields and the treatment by Henneaux and Teitelboim \cite{Henneaux}  for Unimodular Gravity, we have reformulated a TDiff invariant theory for a scalar field $\psi$ in a way in which the invariance of the theory under general Diffeomorphisms is recovered at the expense of introducing an additional vector field $T^\mu$. We have highlighted that the two approaches are indeed equivalent ways of dealing with the same problem. In the TDiff approach, the main tool was the study of the EMT conservation. Indeed, due to the lack of Diff invariance, it was not an automatic consequence of the symmetry, but rather a consistency condition which should be satisfied by the solutions of the theory. On the other hand, in the covariantized approach the symmetry is restored and the EMT is once again trivially conserved. In this case, then, the additional information is provided by the EoM of the newly incorporated vector field $T^\mu$.

We have not only shown the equivalence of the two approaches at the level of general actions, but we have also recovered the results in the potential and kinetic regimes. Furthermore, in the kinetic regime the covariantized treatment has shed light on the conditions to be satisfied by the coupling function to obtain stable adiabatic fluids. We have also discussed how simple models in the kinetic regime can be found following both approaches. Apart from models leading to a constant equation of state parameter, which appear as a power-law for the coupling function of the corresponding approach, a natural model in one approach will not necessarily be simple in the other.

On the other hand, we have shown that the covariantized treatment also yields a novel and important result which was too difficult to obtain in the TDiff approach, namely, the general constraint on the metric whenever both kinetic and potential terms are present. The study of this general result reveals that we may describe our matter content as a non-adiabatic perfect fluid. We have calculated the effective speed of sound of this fluid and found conditions on the allowed coupling functions in order to have physically viable models. Then, we have easily translated all of these new results to the TDiff framework as well. Now, although the most general situation is non-adiabatic, we can find some particular subcases which give adiabatic perturbations. We have found a total of six adiabatic models and performed a detailed analysis of their physical implications. One of them was already known (in particular, the shift-symmetric model was known to be adiabatic), but we have been able to find and study the other models thanks to the simplified treatment provided by the covariantized approach. Finally, we have also discussed a particular solution to a family of TDiff theories which yields the same results as GR. Moreover, we have studied the stability of this solution, finding that it is unstable in the potential regime, while in the kinetic regime the stability depends on the expansion.

Overall, we conclude that the covariantized treatment carried out in this work is a useful tool for several reasons. From the fundamental point of view, the restoration of a symmetry via the introduction of additional quantities can be useful in what respects discerning which are the truly dynamical components of a theory. This shall all be further pursued in future work. From the more practical point of view, the covariantized approach has proven to be faster for particular calculations and, most importantly, it has also revealed a previously unknown result of general character (the implications of which have been thoroughly studied). Nevertheless, we have seen how the more natural models in one approach are not necessarily those that one would consider in the other approach in the first place.  So, one could be missing interesting phenomenology by focusing only in one of the approaches. Moreover, it is intriguing to consider what other problems and analyses, perhaps difficult in one approach, might the other approach help simplify. In addition, having two separate yet equivalent ways of tackling the same problem provides one with a consistency check of sorts, in the sense that a result obtained via one approach may be compared with that obtained through the other. For all of these reasons, we think that combining both approaches will allow us to better understand different aspects of the theory.

\acknowledgments
The authors would like to thank an anonymous referee of reference \cite{Maroto} for useful comments and suggestions which sparked the analysis in this work,  as well as an anonymous referee of the present paper for motivating us to develop its content further. We also wish to thank Javier Rubio for useful discussions and for bringing reference \cite{Blas:2011ac} to our attention. This work has been supported by the MICIN (Spain) project PID2022-138263NB-I00 (AEI/FEDER, UE). DJG also acknowledges financial support from the Ayudas de Doctorado IPARCOS-UCM/2023.

\appendix

\section{Different covariantizations}\label{appendix: Different covariantizations}
As we mentioned in section \ref{section: Covariantized action}, the way of restoring the Diff symmetry is not unique. We have chosen to do so with a vector field $T^\mu$ following Henneaux and Teitelboim \cite{Henneaux} because, in this way, locality is preserved in the covariantized action, but we could very well follow the route taken by Blas et al. in reference \cite{Blas:2011ac}, where they directly consider introducing a new scalar field. There are some differences, however, the discussion of which has been reserved for the present Appendix.

We begin by noting that we can add an arbitrary constant $C_0$ to our TDiff Lagrangian without affecting the EoM of the theory, i.e. we may work equally well with
\begin{equation}\label{eq: TDiff action with C0}
    S_\text{TDiff} = S_\text{EH} + \int d^4x\, \big[ f_k(g) X - f_v(g) V + C_0\big] \,.
\end{equation}
Let us now covariantize this action in the way we presented in section \ref{section: Covariantized action}. We obtain the following expression:
\begin{equation}\label{eq: Diff action with C0}
    S_\text{Diff} = S_\text{EH} + \int d^4x\, \sqrt{g}\big[ H_k(Y) X - H_v(Y) V + C_0 Y \big] \,,
\end{equation}
where we still have not specified how the scalar density $\bar{\mu}$ (equivalently, the combination $Y=\bar{\mu}/\sqrt{g}$) is related to the new field. Let us now do precisely that: we shall introduce a new scalar field $\sigma$ to the theory, and a simple scalar density which we may construct with it is
\begin{equation}
    \bar{\mu} = \sigma \sqrt{g} \,,
\end{equation}
which transforms as wanted. As a result of this choice, we have
\begin{equation}
    Y = \sigma \,,
\end{equation}
and so our Diff action becomes
\begin{equation}
    S_\text{Diff} = S_\text{EH} + \int d^4x\, \sqrt{g}\big[ H_k(\sigma) X - H_v(\sigma) V + C_0 \sigma \big] \,,
\end{equation}
which is a theory for $g_{\mu\nu}$ and two scalar fields ($\psi$ and $\sigma$). If we now take variations with respect to the new scalar field $\sigma$ we find that its EoM is
\begin{equation}\label{eq: EoM sigma}
    H_k'(\sigma) X - H_v'(\sigma) V + C_0 = 0 \,,
\end{equation}
which is practically identical in form to the integrated EoM \eqref{eq: EoM Tmu solutions} for $T^\mu$.

Despite their formal similarities, there are a couple of subtle differences which we wish to stress at this point. The first is that, even though we started by including a superfluous constant in the TDiff action (where it had no effect on the EoM of the theory), in the covariantized approach this arbitrary constant is no longer superfluous as it explicitly enters the EoM of the theory, as shown by equation \eqref{eq: EoM sigma}. As explained in reference \cite{Blas:2011ac}, there is no more a one-to-one correspondence between the EoM in the TDiff approach and those in the covariantized approach: those obtained from the TDiff action \eqref{eq: TDiff action with C0} correspond to a whole family of the ones obtained from the covariantized action \eqref{eq: Diff action with C0}. The constant $C_0$ represents a global degree of freedom which appears explicitly in the action, not fixed in any way by the fields in the theory (recall that, when we included a vector $T^\mu$, the constant $c_\rho$ could be fixed with the initial conditions on the fields and their derivatives). The non-locality in the covariantized action implied by this new global degree of freedom is what we were trying to avoid when we introduced a vector field. Beyond this fact, the second difference we wish to highlight is that the EoM \eqref{eq: EoM sigma} for $\sigma$ is not really dynamical (as opposed to \eqref{eq: EoM Tmu solutions}, where derivatives of $T^\mu$ appear). As a result, the newly included scalar field $\sigma$ would be a spectator field.

Finally, we stress that although the first difference (i.e. having a new global degree of freedom) is a general feature of covariantizing with a scalar field, the second difference (i.e. $\sigma$ being a non-dynamical, spectator field) is a result of our particular field theory. Indeed, recall that we are not breaking the symmetry in the gravitational sector but only in the matter sector, where the determinant appears in the volume element. If we, for instance, decided to break the symmetry in the gravitational sector by including a kinetic term of the form $\partial^\mu g \partial_\mu g$ then, after covariantizing, the scalar field $\sigma$ would have dynamics. All of these points are studied with greater detail in reference \cite{Blas:2011ac}.

\section{Covariantized EMT}\label{appendix: Covariantized EMT}
We include in this Appendix a couple of calculations regarding the covariantized EMT, which are not needed to follow the main thread of the discussion, but which could come in handy in order to clear up some of the computations. The first one has to do with explicitly obtaining expression \eqref{eq: covariantized EMT} for the covariantized EMT, and the second one with its (trivial) conservation.

Let us begin by obtaining it. As we know from definition \eqref{eq: def EMT}, we must find the functional derivative of the matter action
\begin{equation}
    S_m = \int d^4x\, \sqrt{g} \left[ H_k(Y) X - H_v(Y) V(\psi)  \right]
\end{equation}
with respect to the metric. To this end, we shall find the variation $\delta S_m$ with respect to the metric, which reads
\begin{equation}\label{eq: delta Sm}
    \begin{split}
        \delta S_m &= \int d^4x\, \bigg\{ \left( H_k X - H_v V  \right) \delta\sqrt{g} + \sqrt{g}\,X\delta H_k \\
        &\qquad + \sqrt{g}\,H_k\delta X - \sqrt{g}\,V\delta H_v - \sqrt{g}\,H_v\delta V \bigg\} \,.
    \end{split}
\end{equation}
The needed variations with respect to the metric are the following:
\begin{subequations}
    \begin{align}
        \delta V &= 0 \,,\\[5pt]
        \delta X &= \frac{1}{2} \partial_\mu\psi \partial_\nu \psi \,\delta g^{\mu\nu} \,,\\[5pt]
        \delta\sqrt{g} &= - \frac{1}{2} \sqrt{g} \, g_{\mu\nu} \,\delta g^{\mu\nu} \,,\\[5pt]
        \delta H &= H' \left[ \frac{1}{2} Y g_{\mu\nu}\,\delta g^{\mu\nu} + \frac{1}{\sqrt{g}}\partial_\alpha\left( T^\alpha \, \delta\sqrt{g} \right) \right] \,. \label{subeq: delta H}
    \end{align}
\end{subequations}
The usage of the first three is simple enough, but the fourth one deserves a bit more care. After substituting these expressions into the variation \eqref{eq: delta Sm} of the matter action, the integrand will have a term of the form
\begin{equation}
    \begin{split}
        \sqrt{g}\,X \delta H_k &= \sqrt{g} \, X H_k' \frac{1}{2} Y g_{\mu\nu}\,\delta g^{\mu\nu} - T^\alpha \delta\sqrt{g} \, \partial_\alpha\left( H_k' X \right)\\
        &\quad + \partial_\alpha\left( T^\alpha H_k' X \delta\sqrt{g} \right)
    \end{split}
\end{equation}
and a term of the form
\begin{equation}
    \begin{split}
        \sqrt{g}\,V \delta H_v &= \sqrt{g} \,V H_v' \frac{1}{2} Y g_{\mu\nu}\,\delta g^{\mu\nu} - T^\alpha \delta\sqrt{g} \, \partial_\alpha\left( H_v' V \right)\\
        &\quad + \partial_\alpha\left( T^\alpha H_v' V \delta\sqrt{g} \right) \,.
    \end{split}
\end{equation}
A couple of steps are now needed for simplification. The first step is to, as usual, discard the boundary terms appearing in the last lines of the above two expressions, i.e. the ones arising from $\partial_\alpha\left( T^\alpha H_k' X \delta\sqrt{g} \right)$ and $\partial_\alpha\left( T^\alpha H_v' V \delta\sqrt{g} \right)$. The second step is to recognize that after performing the combination $\sqrt{g}\,X \delta H_k - \sqrt{g}\,V \delta H_v$ we will obtain (among others) a term of the form
\begin{equation}
    T^\alpha \delta\sqrt{g} \, \partial_\alpha\left( H_v' V - H_k' X \right) = 0 \,,
\end{equation}
which vanishes by virtue of the EoM \eqref{eq: EoM Tmu} for the vector field $T^\mu$. Carefully substituting and taking everything into account, we finally obtain
\begin{equation}
    \begin{split}
        T_{\mu\nu} &= H_k(Y) \partial_\mu \psi \partial_\nu \psi - \left[ H_k(Y) X - H_v(Y) V \right] g_{\mu\nu} \\[5pt]
        &\qquad + Y \left[ H_k'(Y) X - H_v'(Y) V \right] g_{\mu\nu} \,,
    \end{split}
\end{equation}
which is precisely the covariantized EMT \eqref{eq: covariantized EMT}.\\

The second comment we wish to make is regarding the conservation of this EMT, which should be trivial given that the theory has Diff symmetry, and it can be explicitly checked that this is so. Proceeding piece by piece, we have that
\begin{subequations}
    \begin{align}
        \begin{split}
            \nabla_\mu\left( H_k \nabla^\mu \psi \nabla^\nu \psi \right) &= \nabla^\nu \psi \nabla_\mu\left(H_k \nabla^\mu\psi\right)\\[5pt]  &\quad + H_k \nabla^\mu\psi \left( \nabla_\mu \nabla^\nu \psi \right) \,,
        \end{split}\\
        \begin{split}
            \nabla_\mu\left[ \left( H_k X - H_v V \right) g^{\mu\nu} \right] &= H_k \nabla^\mu\psi \left( \nabla_\mu \nabla^\nu \psi \right)\\[5pt]
            &\quad + \nabla^\nu Y \left( H_k' X - H_v' V  \right)\\[5pt]
            &\quad - H_v V' \nabla^\nu\psi \,,
        \end{split}\\
        \begin{split}
            \nabla_\mu\left[ Y \left( H_k' X - H_v' V \right) g^{\mu\nu} \right] &= \nabla^\nu Y \left( H_k' X - H_v' V  \right)\\[5pt]
            &\quad + Y \nabla^\nu\left( H_k' X - H_v' V  \right) \,,
        \end{split}
    \end{align}
\end{subequations}
Joining now all the pieces together we can see some very convenient cancellations, and also that
\begin{equation}
    \begin{split}
        0=\nabla_\mu T^{\mu\nu} &= \nabla^\nu \psi \underbrace{\left[ \nabla_\mu\left(H_k \nabla^\mu \psi\right) + H_v V' \right]}_{=\,0\,\,\text{by $\psi$ EoM \eqref{eq: covar EoM psi}}} \\[5pt]
        &\quad + Y \underbrace{ \nabla^\nu\left( H_k' X - H_v' V  \right)}_{=\,0\,\,\text{by $T^\mu$ EoM \eqref{eq: EoM Tmu}}} = 0 \,,
    \end{split}
\end{equation}
i.e. we indeed obtain that the conservation of the EMT is trivially satisfied on the solutions to the EoM of the theory, as we expected.

\bibliography{bibliography}

\begin{thebibliography}{23}%
\makeatletter
\providecommand \@ifxundefined [1]{%
 \@ifx{#1\undefined}
}%
\providecommand \@ifnum [1]{%
 \ifnum #1\expandafter \@firstoftwo
 \else \expandafter \@secondoftwo
 \fi
}%
\providecommand \@ifx [1]{%
 \ifx #1\expandafter \@firstoftwo
 \else \expandafter \@secondoftwo
 \fi
}%
\providecommand \natexlab [1]{#1}%
\providecommand \enquote  [1]{``#1''}%
\providecommand \bibnamefont  [1]{#1}%
\providecommand \bibfnamefont [1]{#1}%
\providecommand \citenamefont [1]{#1}%
\providecommand \href@noop [0]{\@secondoftwo}%
\providecommand \href [0]{\begingroup \@sanitize@url \@href}%
\providecommand \@href[1]{\@@startlink{#1}\@@href}%
\providecommand \@@href[1]{\endgroup#1\@@endlink}%
\providecommand \@sanitize@url [0]{\catcode `\\12\catcode `\$12\catcode `\&12\catcode `\#12\catcode `\^12\catcode `\_12\catcode `\%12\relax}%
\providecommand \@@startlink[1]{}%
\providecommand \@@endlink[0]{}%
\providecommand \url  [0]{\begingroup\@sanitize@url \@url }%
\providecommand \@url [1]{\endgroup\@href {#1}{\urlprefix }}%
\providecommand \urlprefix  [0]{URL }%
\providecommand \Eprint [0]{\href }%
\providecommand \doibase [0]{https://doi.org/}%
\providecommand \selectlanguage [0]{\@gobble}%
\providecommand \bibinfo  [0]{\@secondoftwo}%
\providecommand \bibfield  [0]{\@secondoftwo}%
\providecommand \translation [1]{[#1]}%
\providecommand \BibitemOpen [0]{}%
\providecommand \bibitemStop [0]{}%
\providecommand \bibitemNoStop [0]{.\EOS\space}%
\providecommand \EOS [0]{\spacefactor3000\relax}%
\providecommand \BibitemShut  [1]{\csname bibitem#1\endcsname}%
\let\auto@bib@innerbib\@empty
\bibitem [{\citenamefont {Akrami}\ \emph {et~al.}(2021)\citenamefont {Akrami} \emph {et~al.}}]{CANTATA}%
  \BibitemOpen
  \bibfield  {author} {\bibinfo {author} {\bibfnamefont {Y.}~\bibnamefont {Akrami}} \emph {et~al.} (\bibinfo {collaboration} {CANTATA}),\ }\href {https://doi.org/10.1007/978-3-030-83715-0} {\emph {\bibinfo {title} {{Modified Gravity and Cosmology}: {An Update by the CANTATA Network}}}}\ (\bibinfo  {publisher} {Springer},\ \bibinfo {year} {2021})\ \Eprint {https://arxiv.org/abs/2105.12582} {arXiv:2105.12582 [gr-qc]} \BibitemShut {NoStop}%
\bibitem [{\citenamefont {Einstein}(1919)}]{Einstein}%
  \BibitemOpen
  \bibfield  {author} {\bibinfo {author} {\bibfnamefont {A.}~\bibnamefont {Einstein}},\ }\bibfield  {title} {\bibinfo {title} {{Spielen Gravitationsfelder im Aufbau der materiellen Elementarteilchen eine wesentliche Rolle?}},\ }\href@noop {} {\bibfield  {journal} {\bibinfo  {journal} {Sitzungsber. Preuss. Akad. Wiss. Berlin (Math. Phys. )}\ }\textbf {\bibinfo {volume} {1919}},\ \bibinfo {pages} {349} (\bibinfo {year} {1919})}\BibitemShut {NoStop}%
\bibitem [{\citenamefont {Carballo-Rubio}\ \emph {et~al.}(2022)\citenamefont {Carballo-Rubio}, \citenamefont {Garay},\ and\ \citenamefont {Garc\'\i{}a-Moreno}}]{Carballo-Rubio}%
  \BibitemOpen
  \bibfield  {author} {\bibinfo {author} {\bibfnamefont {R.}~\bibnamefont {Carballo-Rubio}}, \bibinfo {author} {\bibfnamefont {L.~J.}\ \bibnamefont {Garay}},\ and\ \bibinfo {author} {\bibfnamefont {G.}~\bibnamefont {Garc\'\i{}a-Moreno}},\ }\bibfield  {title} {\bibinfo {title} {{Unimodular gravity vs general relativity: a status report}},\ }\href {https://doi.org/10.1088/1361-6382/aca386} {\bibfield  {journal} {\bibinfo  {journal} {Class. Quant. Grav.}\ }\textbf {\bibinfo {volume} {39}},\ \bibinfo {pages} {243001} (\bibinfo {year} {2022})},\ \Eprint {https://arxiv.org/abs/2207.08499} {arXiv:2207.08499 [gr-qc]} \BibitemShut {NoStop}%
\bibitem [{\citenamefont {Ellis}\ \emph {et~al.}(2011)\citenamefont {Ellis}, \citenamefont {van Elst}, \citenamefont {Murugan},\ and\ \citenamefont {Uzan}}]{Ellis}%
  \BibitemOpen
  \bibfield  {author} {\bibinfo {author} {\bibfnamefont {G.~F.~R.}\ \bibnamefont {Ellis}}, \bibinfo {author} {\bibfnamefont {H.}~\bibnamefont {van Elst}}, \bibinfo {author} {\bibfnamefont {J.}~\bibnamefont {Murugan}},\ and\ \bibinfo {author} {\bibfnamefont {J.-P.}\ \bibnamefont {Uzan}},\ }\bibfield  {title} {\bibinfo {title} {{On the Trace-Free Einstein Equations as a Viable Alternative to General Relativity}},\ }\href {https://doi.org/10.1088/0264-9381/28/22/225007} {\bibfield  {journal} {\bibinfo  {journal} {Class. Quant. Grav.}\ }\textbf {\bibinfo {volume} {28}},\ \bibinfo {pages} {225007} (\bibinfo {year} {2011})},\ \Eprint {https://arxiv.org/abs/1008.1196} {arXiv:1008.1196 [gr-qc]} \BibitemShut {NoStop}%
\bibitem [{\citenamefont {Maroto}(2024)}]{Maroto}%
  \BibitemOpen
  \bibfield  {author} {\bibinfo {author} {\bibfnamefont {A.~L.}\ \bibnamefont {Maroto}},\ }\bibfield  {title} {\bibinfo {title} {{TDiff invariant field theories for cosmology}},\ }\href {https://doi.org/10.1088/1475-7516/2024/04/037} {\bibfield  {journal} {\bibinfo  {journal} {JCAP}\ }\textbf {\bibinfo {volume} {04}},\ \bibinfo {pages} {037}},\ \Eprint {https://arxiv.org/abs/2301.05713} {arXiv:2301.05713 [gr-qc]} \BibitemShut {NoStop}%
\bibitem [{\citenamefont {Bello-Morales}\ and\ \citenamefont {Maroto}(2024)}]{Bello-Morales}%
  \BibitemOpen
  \bibfield  {author} {\bibinfo {author} {\bibfnamefont {A.~G.}\ \bibnamefont {Bello-Morales}}\ and\ \bibinfo {author} {\bibfnamefont {A.~L.}\ \bibnamefont {Maroto}},\ }\bibfield  {title} {\bibinfo {title} {{Cosmology in gravity models with broken diffeomorphisms}},\ }\href {https://doi.org/10.1103/PhysRevD.109.043506} {\bibfield  {journal} {\bibinfo  {journal} {Phys. Rev. D}\ }\textbf {\bibinfo {volume} {109}},\ \bibinfo {pages} {043506} (\bibinfo {year} {2024})},\ \Eprint {https://arxiv.org/abs/2308.00635} {arXiv:2308.00635 [gr-qc]} \BibitemShut {NoStop}%
\bibitem [{\citenamefont {Pirogov}(2010)}]{Pirogov:2009}%
  \BibitemOpen
  \bibfield  {author} {\bibinfo {author} {\bibfnamefont {Y.~F.}\ \bibnamefont {Pirogov}},\ }\bibfield  {title} {\bibinfo {title} {{Unimodular metagravity versus General Relativity with a scalar field}},\ }\href {https://doi.org/10.1134/S1063778810010151} {\bibfield  {journal} {\bibinfo  {journal} {Phys. Atom. Nucl.}\ }\textbf {\bibinfo {volume} {73}},\ \bibinfo {pages} {134} (\bibinfo {year} {2010})},\ \Eprint {https://arxiv.org/abs/0903.2018} {arXiv:0903.2018 [gr-qc]} \BibitemShut {NoStop}%
\bibitem [{\citenamefont {Pirogov}(2012)}]{Pirogov:2011}%
  \BibitemOpen
  \bibfield  {author} {\bibinfo {author} {\bibfnamefont {Y.~F.}\ \bibnamefont {Pirogov}},\ }\bibfield  {title} {\bibinfo {title} {{Unimodular bimode gravity and the coherent scalar-graviton field as galaxy dark matter}},\ }\href {https://doi.org/10.1140/epjc/s10052-012-2017-y} {\bibfield  {journal} {\bibinfo  {journal} {Eur. Phys. J. C}\ }\textbf {\bibinfo {volume} {72}},\ \bibinfo {pages} {2017} (\bibinfo {year} {2012})},\ \Eprint {https://arxiv.org/abs/1111.1437} {arXiv:1111.1437 [gr-qc]} \BibitemShut {NoStop}%
\bibitem [{\citenamefont {Pirogov}(2006)}]{Pirogov:2005}%
  \BibitemOpen
  \bibfield  {author} {\bibinfo {author} {\bibfnamefont {Y.~F.}\ \bibnamefont {Pirogov}},\ }\bibfield  {title} {\bibinfo {title} {{General covariance violation and the gravitational dark matter. I. Scalar graviton}},\ }\href {https://doi.org/10.1134/S1063778806080102} {\bibfield  {journal} {\bibinfo  {journal} {Phys. Atom. Nucl.}\ }\textbf {\bibinfo {volume} {69}},\ \bibinfo {pages} {1338} (\bibinfo {year} {2006})},\ \Eprint {https://arxiv.org/abs/gr-qc/0505031} {arXiv:gr-qc/0505031} \BibitemShut {NoStop}%
\bibitem [{\citenamefont {Pirogov}(2016)}]{Pirogov:2015}%
  \BibitemOpen
  \bibfield  {author} {\bibinfo {author} {\bibfnamefont {Y.~F.}\ \bibnamefont {Pirogov}},\ }\bibfield  {title} {\bibinfo {title} {{Quartet-metric general relativity: scalar graviton, dark matter and dark energy}},\ }\href {https://doi.org/10.1140/epjc/s10052-016-3973-4} {\bibfield  {journal} {\bibinfo  {journal} {Eur. Phys. J. C}\ }\textbf {\bibinfo {volume} {76}},\ \bibinfo {pages} {215} (\bibinfo {year} {2016})},\ \Eprint {https://arxiv.org/abs/1511.04742} {arXiv:1511.04742 [gr-qc]} \BibitemShut {NoStop}%
\bibitem [{\citenamefont {Alvarez}\ \emph {et~al.}(2009)\citenamefont {Alvarez}, \citenamefont {Faedo},\ and\ \citenamefont {Lopez-Villarejo}}]{Alvarez}%
  \BibitemOpen
  \bibfield  {author} {\bibinfo {author} {\bibfnamefont {E.}~\bibnamefont {Alvarez}}, \bibinfo {author} {\bibfnamefont {A.~F.}\ \bibnamefont {Faedo}},\ and\ \bibinfo {author} {\bibfnamefont {J.~J.}\ \bibnamefont {Lopez-Villarejo}},\ }\bibfield  {title} {\bibinfo {title} {{Transverse gravity versus observations}},\ }\href {https://doi.org/10.1088/1475-7516/2009/07/002} {\bibfield  {journal} {\bibinfo  {journal} {JCAP}\ }\textbf {\bibinfo {volume} {07}}\bibfield  {number} {\bibinfo  {number} { (2009)},\ \bibinfo {pages} {002}},\ }\Eprint {https://arxiv.org/abs/0904.3298} {arXiv:0904.3298 [hep-th]} \BibitemShut {NoStop}%
\bibitem [{\citenamefont {Jaramillo-Garrido}\ \emph {et~al.}(2024)\citenamefont {Jaramillo-Garrido}, \citenamefont {Maroto},\ and\ \citenamefont {Mart\'in-Moruno}}]{Jaramillo-Garrido}%
  \BibitemOpen
  \bibfield  {author} {\bibinfo {author} {\bibfnamefont {D.}~\bibnamefont {Jaramillo-Garrido}}, \bibinfo {author} {\bibfnamefont {A.~L.}\ \bibnamefont {Maroto}},\ and\ \bibinfo {author} {\bibfnamefont {P.}~\bibnamefont {Mart\'in-Moruno}},\ }\bibfield  {title} {\bibinfo {title} {{TD}iff in the dark: gravity with a scalar field invariant under transverse diffeomorphisms},\ }\href {https://doi.org/10.1007/JHEP03(2024)084} {\bibfield  {journal} {\bibinfo  {journal} {JHEP}\ }\textbf {\bibinfo {volume} {03}}\bibfield  {number} {\bibinfo  {number} { (2024)},\ \bibinfo {pages} {084}},\ }\Eprint {https://arxiv.org/abs/2307.14861} {arXiv:2307.14861 [gr-qc]} \BibitemShut {NoStop}%
\bibitem [{\citenamefont {Alonso-L\'opez}\ \emph {et~al.}(2024)\citenamefont {Alonso-L\'opez}, \citenamefont {de~Cruz~P\'erez},\ and\ \citenamefont {Maroto}}]{Alonso-Lopez}%
  \BibitemOpen
  \bibfield  {author} {\bibinfo {author} {\bibfnamefont {D.}~\bibnamefont {Alonso-L\'opez}}, \bibinfo {author} {\bibfnamefont {J.}~\bibnamefont {de~Cruz~P\'erez}},\ and\ \bibinfo {author} {\bibfnamefont {A.~L.}\ \bibnamefont {Maroto}},\ }\bibfield  {title} {\bibinfo {title} {{Unified transverse diffeomorphism invariant field theory for the dark sector}},\ }\href {https://doi.org/10.1103/PhysRevD.109.023537} {\bibfield  {journal} {\bibinfo  {journal} {Phys. Rev. D}\ }\textbf {\bibinfo {volume} {109}},\ \bibinfo {pages} {023537} (\bibinfo {year} {2024})},\ \Eprint {https://arxiv.org/abs/2311.16836} {arXiv:2311.16836 [astro-ph.CO]} \BibitemShut {NoStop}%
\bibitem [{\citenamefont {Afonso}\ \emph {et~al.}(2019)\citenamefont {Afonso}, \citenamefont {Olmo}, \citenamefont {Orazi},\ and\ \citenamefont {Rubiera-Garcia}}]{Afonso:2018hyj}%
  \BibitemOpen
  \bibfield  {author} {\bibinfo {author} {\bibfnamefont {V.~I.}\ \bibnamefont {Afonso}}, \bibinfo {author} {\bibfnamefont {G.~J.}\ \bibnamefont {Olmo}}, \bibinfo {author} {\bibfnamefont {E.}~\bibnamefont {Orazi}},\ and\ \bibinfo {author} {\bibfnamefont {D.}~\bibnamefont {Rubiera-Garcia}},\ }\bibfield  {title} {\bibinfo {title} {{Correspondence between modified gravity and general relativity with scalar fields}},\ }\href {https://doi.org/10.1103/PhysRevD.99.044040} {\bibfield  {journal} {\bibinfo  {journal} {Phys. Rev. D}\ }\textbf {\bibinfo {volume} {99}},\ \bibinfo {pages} {044040} (\bibinfo {year} {2019})},\ \Eprint {https://arxiv.org/abs/1810.04239} {arXiv:1810.04239 [gr-qc]} \BibitemShut {NoStop}%
\bibitem [{\citenamefont {Stueckelberg}(1938)}]{Stueckelberg}%
  \BibitemOpen
  \bibfield  {author} {\bibinfo {author} {\bibfnamefont {E.~C.~G.}\ \bibnamefont {Stueckelberg}},\ }\bibfield  {title} {\bibinfo {title} {{Interaction energy in electrodynamics and in the field theory of nuclear forces}},\ }\href {https://doi.org/10.5169/seals-110852} {\bibfield  {journal} {\bibinfo  {journal} {Helv. Phys. Acta}\ }\textbf {\bibinfo {volume} {11}},\ \bibinfo {pages} {225} (\bibinfo {year} {1938})}\BibitemShut {NoStop}%
\bibitem [{\citenamefont {Ruegg}\ and\ \citenamefont {Ruiz-Altaba}(2004)}]{Ruegg}%
  \BibitemOpen
  \bibfield  {author} {\bibinfo {author} {\bibfnamefont {H.}~\bibnamefont {Ruegg}}\ and\ \bibinfo {author} {\bibfnamefont {M.}~\bibnamefont {Ruiz-Altaba}},\ }\bibfield  {title} {\bibinfo {title} {{The Stueckelberg field}},\ }\href {https://doi.org/10.1142/S0217751X04019755} {\bibfield  {journal} {\bibinfo  {journal} {Int. J. Mod. Phys. A}\ }\textbf {\bibinfo {volume} {19}},\ \bibinfo {pages} {3265} (\bibinfo {year} {2004})},\ \Eprint {https://arxiv.org/abs/hep-th/0304245} {arXiv:hep-th/0304245} \BibitemShut {NoStop}%
\bibitem [{\citenamefont {Henneaux}\ and\ \citenamefont {Teitelboim}(1989)}]{Henneaux}%
  \BibitemOpen
  \bibfield  {author} {\bibinfo {author} {\bibfnamefont {M.}~\bibnamefont {Henneaux}}\ and\ \bibinfo {author} {\bibfnamefont {C.}~\bibnamefont {Teitelboim}},\ }\bibfield  {title} {\bibinfo {title} {{The Cosmological Constant and General Covariance}},\ }\href {https://doi.org/10.1016/0370-2693(89)91251-3} {\bibfield  {journal} {\bibinfo  {journal} {Phys. Lett. B}\ }\textbf {\bibinfo {volume} {222}},\ \bibinfo {pages} {195} (\bibinfo {year} {1989})}\BibitemShut {NoStop}%
\bibitem [{\citenamefont {Kuchar}(1991)}]{Kuchar}%
  \BibitemOpen
  \bibfield  {author} {\bibinfo {author} {\bibfnamefont {K.~V.}\ \bibnamefont {Kuchar}},\ }\bibfield  {title} {\bibinfo {title} {{Does an unspecified cosmological constant solve the problem of time in quantum gravity?}},\ }\href {https://doi.org/10.1103/PhysRevD.43.3332} {\bibfield  {journal} {\bibinfo  {journal} {Phys. Rev. D}\ }\textbf {\bibinfo {volume} {43}},\ \bibinfo {pages} {3332} (\bibinfo {year} {1991})}\BibitemShut {NoStop}%
\bibitem [{\citenamefont {de~Rham}\ \emph {et~al.}(2012)\citenamefont {de~Rham}, \citenamefont {Gabadadze},\ and\ \citenamefont {Tolley}}]{deRham:2011rn}%
  \BibitemOpen
  \bibfield  {author} {\bibinfo {author} {\bibfnamefont {C.}~\bibnamefont {de~Rham}}, \bibinfo {author} {\bibfnamefont {G.}~\bibnamefont {Gabadadze}},\ and\ \bibinfo {author} {\bibfnamefont {A.~J.}\ \bibnamefont {Tolley}},\ }\bibfield  {title} {\bibinfo {title} {{Ghost free Massive Gravity in the St\"uckelberg language}},\ }\href {https://doi.org/10.1016/j.physletb.2012.03.081} {\bibfield  {journal} {\bibinfo  {journal} {Phys. Lett. B}\ }\textbf {\bibinfo {volume} {711}},\ \bibinfo {pages} {190} (\bibinfo {year} {2012})},\ \Eprint {https://arxiv.org/abs/1107.3820} {arXiv:1107.3820 [hep-th]} \BibitemShut {NoStop}%
\bibitem [{\citenamefont {Poisson}(2004)}]{Poisson}%
  \BibitemOpen
  \bibfield  {author} {\bibinfo {author} {\bibfnamefont {E.}~\bibnamefont {Poisson}},\ }\href {https://doi.org/10.1017/CBO9780511606601} {\emph {\bibinfo {title} {A {Relativist's} {Toolkit}: {The} {Mathematics} of {Black}-{Hole} {Mechanics}}}}\ (\bibinfo  {publisher} {CUP},\ \bibinfo {address} {Cambridge},\ \bibinfo {year} {2004})\BibitemShut {NoStop}%
\bibitem [{\citenamefont {Blas}\ \emph {et~al.}(2011)\citenamefont {Blas}, \citenamefont {Shaposhnikov},\ and\ \citenamefont {Zenhausern}}]{Blas:2011ac}%
  \BibitemOpen
  \bibfield  {author} {\bibinfo {author} {\bibfnamefont {D.}~\bibnamefont {Blas}}, \bibinfo {author} {\bibfnamefont {M.}~\bibnamefont {Shaposhnikov}},\ and\ \bibinfo {author} {\bibfnamefont {D.}~\bibnamefont {Zenhausern}},\ }\bibfield  {title} {\bibinfo {title} {Scale-invariant alternatives to general relativity},\ }\href {https://doi.org/10.1103/PhysRevD.84.044001} {\bibfield  {journal} {\bibinfo  {journal} {Phys. Rev. D}\ }\textbf {\bibinfo {volume} {84}},\ \bibinfo {pages} {044001} (\bibinfo {year} {2011})},\ \Eprint {https://arxiv.org/abs/1104.1392} {arXiv:1104.1392 [hep-th]} \BibitemShut {NoStop}%
\bibitem [{\citenamefont {Hu}(1998)}]{Hu:1998kj}%
  \BibitemOpen
  \bibfield  {author} {\bibinfo {author} {\bibfnamefont {W.}~\bibnamefont {Hu}},\ }\bibfield  {title} {\bibinfo {title} {Structure formation with generalized dark matter},\ }\href {https://doi.org/10.1086/306274} {\bibfield  {journal} {\bibinfo  {journal} {Astrophys. J.}\ }\textbf {\bibinfo {volume} {506}},\ \bibinfo {pages} {485} (\bibinfo {year} {1998})},\ \Eprint {https://arxiv.org/abs/astro-ph/9801234} {arXiv:astro-ph/9801234} \BibitemShut {NoStop}%
\bibitem [{\citenamefont {Gordon}\ and\ \citenamefont {Hu}(2004)}]{Gordon:2004ez}%
  \BibitemOpen
  \bibfield  {author} {\bibinfo {author} {\bibfnamefont {C.}~\bibnamefont {Gordon}}\ and\ \bibinfo {author} {\bibfnamefont {W.}~\bibnamefont {Hu}},\ }\bibfield  {title} {\bibinfo {title} {A low cmb quadrupole from dark energy isocurvature perturbations},\ }\href {https://doi.org/10.1103/PhysRevD.70.083003} {\bibfield  {journal} {\bibinfo  {journal} {Phys. Rev. D}\ }\textbf {\bibinfo {volume} {70}},\ \bibinfo {pages} {083003} (\bibinfo {year} {2004})},\ \Eprint {https://arxiv.org/abs/astro-ph/0406496} {arXiv:astro-ph/0406496} \BibitemShut {NoStop}%
\end{thebibliography}%

\end{document}